\documentclass[preprint,journal]{vgtc}       

\ifpdf
  \pdfoutput=1\relax                   
  \usepackage{graphicx}                
  \DeclareGraphicsExtensions{.pdf,.png,.jpg,.jpeg} 
\else
  \usepackage{graphicx}                
  \DeclareGraphicsExtensions{.eps}     
\fi%

\graphicspath{{figures/}{pictures/}{images/}{./}} 

\usepackage{microtype}                 
\PassOptionsToPackage{warn}{textcomp}  
\usepackage{textcomp}                  
\usepackage{mathptmx}                  
\usepackage{times}                     
\usepackage{cite}                      
\usepackage{tabu}                      
\usepackage{booktabs}                  
\usepackage{bbm}
\usepackage{subfigure}

\usepackage{amsmath}
\usepackage{mathtools}
\usepackage{amsfonts}
\usepackage{verbatim}
\usepackage{placeins}
\newcommand{\Rspace}        {{\mathbb R}}
\newcommand{\fix}[1]{{\textcolor{black}{#1}}}
\newcommand\numberthis{\addtocounter{equation}{1}\tag{\theequation}}

\preprinttext{To appear in IEEE Transactions on Visualization and Computer Graphics.}

\vgtccategory{Research}
\vgtcpapertype{Area 5 - Data Transformations}

\title{Fiber Uncertainty Visualization for Bivariate Data With Parametric and Nonparametric Noise Models}

\author{Tushar M. Athawale, Chris R. Johnson, Sudhanshu Sane, and David Pugmire}
\authorfooter{
\item
Tushar M. Athawale and David Pugmire are with Oak Ridge National Laboratory. E-mail: \{athawaletm $\vert$ pugmire\}@ornl.gov.
\item
Chris R. Johnson is with Scientific Computing \& Imaging (SCI) Institute at the University of Utah. E-mail: crj@sci.utah.edu.
\item
Sudhanshu Sane is with Luminary Cloud, Inc. E-mail: sudhanshu.sane@gmail.com.

This manuscript has been authored by UT-Battelle, LLC under Contract No. DE-AC05-00OR22725 with the U.S. Department of Energy.  The publisher, by accepting the article for publication, acknowledges that the U.S. Government retains a non-exclusive, paid up, irrevocable, world-wide license to publish or reproduce the published form of the manuscript, or allow others to do so, for U.S. Government purposes. The DOE will provide public access to these results in accordance with the DOE Public Access Plan (\url{http://energy.gov/downloads/doe-public-access-plan}).
}

\shortauthortitle{Biv \MakeLowercase{\textit{et al.}}: Global Illumination for Fun and Profit}

\abstract{Visualization and analysis of multivariate data and their uncertainty are top research challenges in data visualization. Constructing fiber surfaces is a popular technique for multivariate data visualization that generalizes the idea of level-set visualization for univariate data to multivariate data. In this paper, we present a statistical framework to quantify positional probabilities of fibers extracted from uncertain bivariate fields. Specifically, \fix{we extend the state-of-the-art Gaussian models of uncertainty for bivariate data to other parametric distributions (e.g., uniform and Epanechnikov) and more general nonparametric probability distributions (e.g., histograms and kernel density estimation)} and derive corresponding spatial probabilities of fibers. In our proposed framework, we leverage Green's theorem for closed-form computation of fiber probabilities when bivariate data are assumed to have {\em independent} parametric and nonparametric noise. Additionally, we present a nonparametric approach combined with numerical integration to study the positional probability of fibers when bivariate data are assumed to have {\em correlated} noise. For uncertainty analysis, we visualize the derived probability volumes for fibers via volume rendering and extracting level sets based on probability thresholds. We present the utility of our proposed techniques via experiments on synthetic and simulation datasets.}


\CCScatlist{ 
 \CCScat{K.6.1}{Management of Computing and Information Systems}%
{Project and People Management}{Life Cycle};
 \CCScat{K.7.m}{The Computing Profession}{Miscellaneous}{Ethics}
}

\vgtcinsertpkg

\begin{document}


\firstsection{Introduction}

\maketitle

In an era in which the growth of and dependence on data is increasing, understanding the inherent uncertainty is critical.
Uncertainty can be introduced in many different ways~\cite{TA:Brodlie:2012:RUDV}.
Experimental and observational data are limited to the accuracy of the sensors.
Accuracy in simulation data is a function of the model, numerical approximation, and parameters (e.g., grid resolution, number of particles) being used.
Additional uncertainty can be introduced by data reduction (e.g., filtering, lossy compression), which is often used to address the challenges of big data. Further, the data are becoming increasingly more complex.
Simulation data are often multivariate, with multiple scalar-, vector-, and/or tensor-quantities simulated over the spatial domain. Effective communication of uncertainty is, therefore, of paramount importance to perform risk-aware science and increase trust in scientific decisions~\cite{TA:Kamal:2021:UQvisSurvey, TA:Zuk:2007:VisUncertaintyReasoning, TA:Johnson:2003:nextStepVisErrors}. Although there have been several advances in multivariate data visualization and analysis \cite{TA:Fuchs:2009:MultivariateSciVis}, relatively few studies have investigated the uncertainty of multivariate data~\cite{TA:Bonneau:2014:StateOftheArtUQ, TA:Potter:2012:UQtaxonomy}. Here, we address the challenge of uncertainty visualization of bivariate scalar fields.

Fiber surfaces as proposed by Carr et al.~\cite{TA:Carr:2015:fiberSurfaces} are a well-known bivariate data visualization technique for exploring correlations among variables. Fiber surfaces generalize the idea of isosurfaces~\cite{TA:Lorensen:1987:MCA} for univariate data to bivariate data. More specifically, given a bivariate function $f: \Rspace^3 \to \Rspace^2$ with a three-dimensional (3D) spatial domain and a two-dimensional (2D) attribute space, fiber surfaces are a preimage of a user-defined {\em fiber surface control polygon} (FSCP) or {\em trait} selected in the 2D attribute space (see~\autoref{sec:fiberSurfaces}). We use the terms FSCP and trait interchangeably. FSCP generalizes the idea of an isovalue for univariate data to higher dimensions. In this paper, we study the uncertainty in fiber positions arising from noise in bivariate data.

Our contributions in this paper are primarily motivated by the recent work by Zheng and Sadlo~\cite{TA:Zheng:2021:uncertainScatterPlots}, who studied the positional uncertainty of fibers for uncertain bivariate fields. They proposed a closed-form uncertainty quantification framework for bivariate data fibers assuming independent/correlated Gaussian (parametric) noise models and rectangular FSCP selection (see~\autoref{sec:stateOftheArtUncertaintyVisFibers}). The probabilities of uncertain fiber positions for rectangular FSCP selection (referred to as {\em uncertain range-fibers}) were visualized with direct volume rendering and were shown to significantly deviate from the fiber positions extracted when assuming no uncertainty. One of the future directions identified by Zheng and Sadlo (see Sect.~7 of~\cite{TA:Zheng:2021:uncertainScatterPlots}) included extending their work to probability distributions other than Gaussian, which we address here.

In this paper, we expand the work by Zheng and Sadlo to other parametric distributions, such as uniform and Epanechnikov, and more general {\em nonparametric} noise distributions (histograms or kernel density estimation [KDE]~\cite{TA:Parzen:1968:ParzenWindow}) for {\em arbitrary} shapes of FSCP. The nonparametric models of uncertainty can provide greater accuracy than the parametric models in the context of visualization~\cite{TA:Athawale:2016:nonparametricIsosurfaces, TA:Kai:2013:nonparametricIsoVis, TA:Athawale:2021:nonparametricDVR} owing to their ability to capture more realistic shapes of underlying probability distributions. Compared with the more restrictive rectangular polygon shapes, FSCP's arbitrary shapes provide an additional flexibility to choose features from the 2D attribute space and visualize their fiber surfaces. For example, Carr et al.~\cite{TA:Carr:2015:fiberSurfaces} and Klacansky et al.~\cite{TA:Klacansky:2017:FiberSurfacesTetrahedralMeshes} studied fiber surface visualizations for arbitrary FSCP shapes in application domains, including chemistry and medical imaging (see~\autoref{sec:fiberSurfaces}).
\vspace{-1mm}
\paragraph{Contributions.} Our contributions in this paper are fourfold. First, we propose a closed-form statistical framework for uncertainty quantification of fibers of bivariate data when noise in data is modeled with {\em independent parametric probability distributions}, such as uniform, Epanechnikov, and Gaussian (\autoref{sec:parametric}). Note that our framework is applicable to FSCP with arbitrary shapes as opposed to being limited to rectangular FSCP~\cite{TA:Zheng:2021:uncertainScatterPlots, TA:Sane2021FCLS}. Our derivations for parametric distributions act as building blocks for our derivations relevant to nonparametric distributions. Second, we present a closed-form statistical framework to derive the positional uncertainty of fibers when noise in bivariate data is modeled with {\em independent nonparametric probability distributions} (e.g., histograms or Parzen window/KDE~\cite{TA:Parzen:1968:ParzenWindow}) (\autoref{sec:nonparametricIndependent}). We present our results for noise kernels, including uniform, Epanechnikov, and Gaussian.
Third, we present a nonparametric statistical framework for uncertainty quantification of bivariate data fibers when noise in variables is assumed to be {\em correlated} (\autoref{sec:nonparametricCorrelated}). We present a closed-form uncertainty quantification of fibers when the noise correlation is estimated with 2D histograms, and a numerical integration technique is proposed to approximate fiber uncertainty when the noise correlation is estimated with bivariate KDE~\cite{TA:1993:Wand:kdeBivariate}. Lastly, we leverage the idea of  {\em vertex-based classification}~\cite{TA:Athawale:2016:nonparametricIsosurfaces} to extract the most probable fiber surface from uncertain bivariate data (\autoref{sec:visualizationTechniques}).

We organize our paper as follows: In \autoref{sec:background}, we define our notation and provide a brief overview of bivariate fiber surfaces (\autoref{sec:fiberSurfaces}) and the state of the art in uncertainty visualization of fibers (\autoref{sec:stateOftheArtUncertaintyVisFibers}). We then discuss relevant prior work in multivariate data and uncertainty visualization in \autoref{sec:relatedWork}. We describe our proposed frameworks for uncertainty quantification of fibers with parametric noise assumptions in \autoref{sec:parametric} and nonparametric noise assumptions in \autoref{sec:nonparametric}. The memory and computational complexities of our proposed statistical techniques are discussed in \autoref{sec:complexity}. We describe our visualization methods in \autoref{sec:visualizationTechniques}. Finally, we present the results of our statistical uncertainty analysis of synthetic and simulation datasets in \autoref{sec:results} and discuss conclusions and potential future directions in \autoref{sec:conclusion}.

\section{Background}\label{sec:background}
\subsection{Fiber Surfaces}\label{sec:fiberSurfaces}
We briefly summarize the fiber surface extraction technique proposed by Carr et al.~\cite{TA:Carr:2015:fiberSurfaces} for bivariate scalar fields. Broadly, for a multivariate function $f: \Rspace^n \to \Rspace^m$, the spatial domain $\mathcal{D} \subset \Rspace^{n}$ is mapped to an attribute space $\mathcal{A} \subset  \Rspace^{m}$. The attribute space $\mathcal{A}$ encodes scientific observations or simulations comprising scalar, vector, and tensor fields. In this paper, we consider bivariate scalar fields with $\mathcal{D} \subset \Rspace^{3}$ and $\mathcal{A} \subset  \Rspace^{2}$. Let $\mathcal{A} = \{\mathcal{A}_1, \mathcal{A}_2\}$ be the attribute space with two attributes or scalar functions $\mathcal{A}_1$ and $\mathcal{A}_2$. A {\em fiber} is then defined as the inverse image of a point $a = (a_1, a_2) \in \mathcal{A}$~\cite{TA:Saeki:2004:singularFibers, TA:Carr:2015:fiberSurfaces, TA:Klacansky:2017:FiberSurfacesTetrahedralMeshes}, and it corresponds to an intersection of isosurfaces~\cite{TA:Lorensen:1987:MCA} for isovalues $\mathcal{A}_1 = a_1$ and  $\mathcal{A}_2 = a_2$.

\begin{figure}[!ht]
 \centering 
 \vspace{-3.8mm}
 \includegraphics[width=0.98\columnwidth]{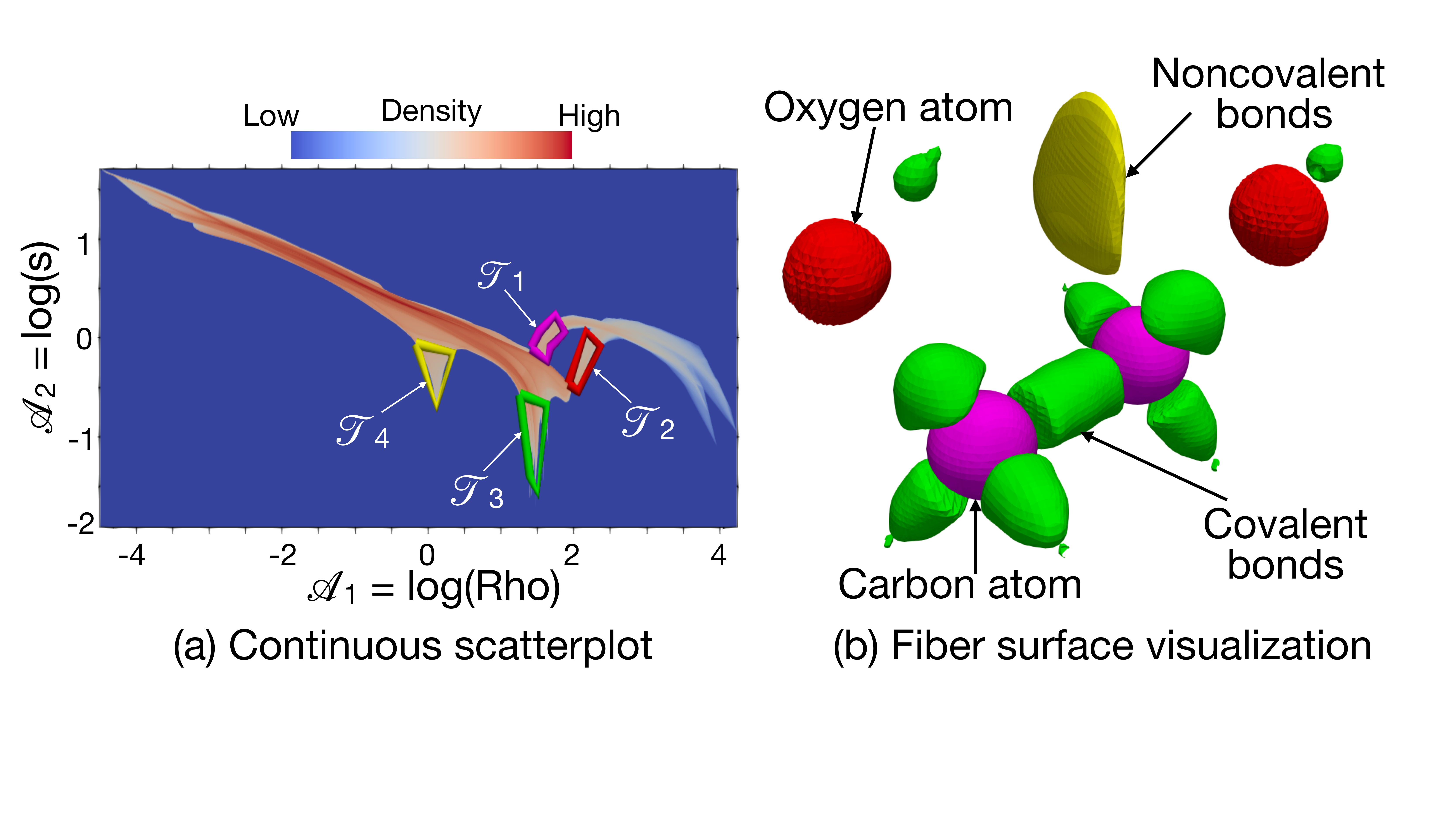}
 \vspace{-10mm}
 \caption{Fiber surface visualization of the ethanediol dataset. (a) Visualization of a \fix{continuous} scatterplot of the electron density (Rho) and reduced gradient (s) at a logarithmic scale. (b) Fiber surfaces for four traits $\{\mathcal{T}_1, \mathcal{T}_2,  \mathcal{T}_3,  \mathcal{T}_4\}$ denoted by polygons in image (a).}
 \label{fig:ethaneDiolFiberSurface}
\end{figure}

The fibers of interest can be specified with FSCP, which we denote as a trait, $\mathcal{T} \subset \mathcal{A}$. Each point on a trait boundary has a corresponding fiber, and such fibers carve out or represent a {\em fiber surface}. Consider the fiber surface example illustrated for the ethandiol dataset in \autoref{fig:ethaneDiolFiberSurface}, which is similar to the one by Carr et al.~\cite{TA:Carr:2015:fiberSurfaces}. \autoref{fig:ethaneDiolFiberSurface}a visualizes a \fix{continuous} scatterplot~\cite{TA:Bachthaler:2008:continuousScatterPlots} of the grid vertex data with the electron density (Rho) as attribute $\mathcal{A}_1$ on the horizontal axis and the reduced gradient (s)~\cite{TA:Johnson:2010:moleculeDataset} as attribute $\mathcal{A}_2$ on the vertical axis. \fix{The continuous scatterplot in all our examples is computed with the topology toolkit~\cite{TA:2018:TFL}}. The four traits {$\{\mathcal{T}_1, \mathcal{T}_2, \mathcal{T}_3,  \mathcal{T}_4\}$} are selected as colored polygons with arbitrary shapes in an attribute space. The respective colored fiber surfaces (i.e., the set of fibers corresponding to points along the polygon boundaries) are visualized in \autoref{fig:ethaneDiolFiberSurface}b.

The process of fiber surface extraction is similar to the process of isosurface extraction, which uses the marching cubes algorithm~\cite{TA:Lorensen:1987:MCA} but with a few tweaks~\cite{TA:Carr:2015:fiberSurfaces}. Initially, each grid vertex of domain $\mathcal{D}$ is classified as {\em interior} or {\em exterior} depending on whether the function values lie inside (interior) or outside (exterior) FSCP. Mathematically, a vertex $v$ is classified as interior if $(\mathcal{A}_1(v), \mathcal{A}_2(v) ) \in \mathcal{T}$; otherwise, it is classified as exterior. Such classification determines the marching cubes topology case within each grid cell for fiber surface reconstruction, and inverse linear interpolation may be applied in the attribute space $\mathcal{A}$ to estimate fiber surface positions on grid edges (please refer to Algorithm 1 of~\cite{TA:Carr:2015:fiberSurfaces}). Feature level sets~\cite{TA:Jankowai:2020:featureLevelSets} generalize the idea of fiber surfaces and extract features based on distance $d$ from trait $\mathcal{T}$. For $d=0$, the feature corresponds to the fiber surface itself. For $d>0$, features present a spatial evolution of fiber surfaces in the vicinity of trait $\mathcal{T}$.
\subsection{Fiber Uncertainty for Rectangular FSCP}\label{sec:stateOftheArtUncertaintyVisFibers}

The uncertainty in multivariate data can result in ambiguity as to whether a vertex should be classified as interior or exterior. The problem of uncertain vertex classification was recently addressed by Zheng and Sadlo~\cite{TA:Zheng:2021:uncertainScatterPlots} and Sane et al.~\cite{TA:Sane2021FCLS}. Zheng and Sadlo proposed a statistical framework to compute the probability of data at point $P \in \mathcal{D}$ being in the interior of a rectangular FSCP.  The computed probabilities were then visualized via direct volume rendering and referred to as uncertain range-fibers. Their analysis assumed the independent/correlated Gaussian (parametric) noise model and a rectangular shape of trait $\mathcal{T}$. 

\begin{figure}[!ht]
 \centering 
 \vspace{-2mm}
 \includegraphics[width=0.78\columnwidth]{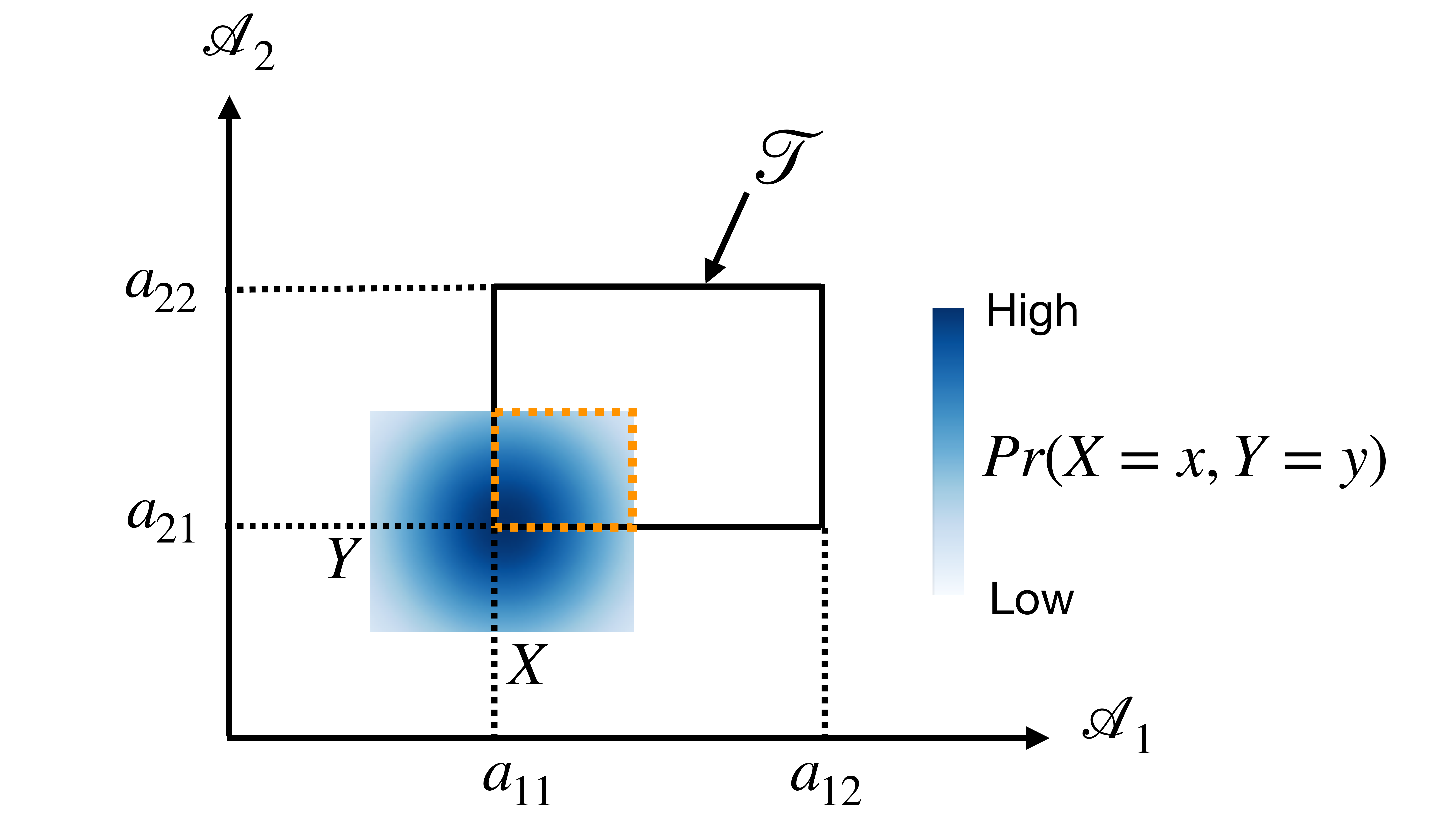}
 \vspace{-2mm}
 \caption{Illustration of the interior probability computation for a single point of a spatial domain when FSCP is rectangular. $X$ and $Y$ denote the uncertain data ranges of a point, with blue denoting the joint probability distribution over the uncertain range. The black rectangle denotes trait $\mathcal{T}$. The probability of a point being in the interior of a fiber surface can be computed by integrating the probability density function (blue) over trait $\mathcal{T}$ or the region enclosed by the dotted orange rectangle.}
 \label{fig:zhengSadloTechnique}
\end{figure}

We illustrate the probability computation process proposed by Zheng and Sadlo in \autoref{fig:zhengSadloTechnique}. Let random variable $\mathcal{U} = (X, Y)$ denote uncertain bivariate data at position $P \in \mathcal{D}$, where $X \subset \mathcal{A}_1$ and $Y \subset \mathcal{A}_2$ are the two random variables that indicate uncertain data ranges for two attributes. Let $\text{\fix{pdf}}_{X}(x)$ and $\text{\fix{pdf}}_{Y}(y)$ denote the probability distributions of random variables $X$ and $Y$, respectively. Let the rectangular trait $\mathcal{T}$ be defined by ranges $(a_{11}, a_{12})$ and  $(a_{21}, a_{22})$ for attributes $\mathcal{A}_1$ and  $\mathcal{A}_2$, respectively. Let $Pr(\mathcal{U} \in \mathcal{T})$ denote the probability of uncertain data at point $P$ being in the interior of a trait, which we refer to as {\em interior probability}. The interior probability (i.e., Pr($\mathcal{U}$ = $(X, Y)$ $\in \mathcal{T}$)) can be computed in two steps: (1)~the joint probability distribution of $X$ and $Y$ (i.e., $\text{\fix{pdf}}_{X,Y}(x,y)$) is computed, and then (2) the joint probability distribution $\text{\fix{pdf}}_{X,Y}(x,y)$ is integrated over trait $\mathcal{T}$. Mathematically, the probability of point $P$ being in the interior of a fiber surface can be expressed as the following double integral over a rectangular trait:

\begin{equation} \label{eq:rectangularIntegration}
Pr(\mathcal{U} = (X,Y) \in \mathcal{T}) = \int_{a_{11}}^{a_{12}} \int_{a_{21}}^{a_{22}}  \text{\fix{pdf}}_{X,Y}(x,y) dx dy
\end{equation}
Zheng and Sadlo~\cite{TA:Zheng:2021:uncertainScatterPlots} proposed a closed-form computation of the double integral in \autoref{eq:rectangularIntegration} when $ \text{\fix{pdf}}_{X, Y} (x,y)$ is Gaussian distributed. Sane et al.~\cite{TA:Sane2021FCLS} proposed confidence-feature level sets for Gaussian distributed data and rectangular FSCP for bivariate fields.
In their technique, they considered confidence intervals of uncertain ranges at a position $P$ (i.e., random variables $X$ and $Y$) to generate respective confidence visualizations of fiber surfaces.

\section{Related Work}\label{sec:relatedWork}
Our proposed methods mainly relate to two important research areas of scientific visualization: multivariate data visualization and uncertainty visualization. Next, we discuss prior work in each of these areas that is relevant to fiber surface visualization. 

\subsection{Multivariate Data Visualization}
 Understanding correlations among variables is one of the main challenges in multivariate data analysis. Scatterplots~\cite{TA:friendly:2005:earlyOriginsOfScatterplot} and parallel coordinate plots~\cite{TA:Inselberg:1991:parallelCoordinates} are fundamental tools that facilitate effective exploration of attribute spaces of multivariate data. Several research efforts have focused on improving challenges associated with scatterplot and parallel-coordinate plot visualization. Continuous versions of scatterplots~\cite{TA:Bachthaler:2008:continuousScatterPlots} and parallel coordinate plots~\cite{TA:Heinrich:2009:continuousParallelCoordinates} have been proposed to avoid sampling artifacts of their traditional discrete versions. Quadri and Rosen~\cite{TA:Quadri:2021:scatterplotClustering} used topological data analysis to identify and explore clusters in scatterplots. Sauber et al.~\cite{TA:Sauber:2006:multifield-Graphs} proposed multifield graphs to study correlations among variables and their correlation strength.

Data variables can be encoded into glyphs~\cite{TA:Ward:2008:glyphsMultifieldDataVis} to visualize a correlation in a physical domain but can suffer from occlusion and cluttering issues. Other effective alternatives for multivariate visualization include brushing of scatterplots or parallel coordinate plots and linking them with the spatial domain for exploration of correlations in a physical space. For example, in \autoref{fig:ethaneDiolFiberSurface}, each trait $\mathcal{T}_i$ indicates brushing in the attribute space, and their respective visualization in physical space corresponds to a fiber surface. Hauser et al.~\cite{TA:2002:Hauser:angularBrushing} introduced angular brushes for exploration of parallel coordinate plots and scatterplots. Kniss et al.~\cite{TA:Kniss:2002:multiDTF} proposed multidimensional transfer functions as a brushing technique for volume rendering of multidimensional data. J\"{a}nicke et al.~\cite{TA:Janicke:2008:dimesionalityReductionBrushingMultiFieldVis} presented an approach for the projection of multivariate data to a 2D space (referred to as \textit{attribute clouds}) followed by clustering and brushing for visualization.

We provide a brief overview of several variants and applications of fiber surfaces for visualizing multivariate data. Klacansky et al.~\cite{TA:Klacansky:2017:FiberSurfacesTetrahedralMeshes} extended the fiber surface extraction technique to tetrahedral meshes and proposed a parallel computational framework for interactive exploration of fiber surfaces. Wu et al.~\cite{TA:Wu:2017:dvrFiberSurfaces} presented a direct volume rendering framework for visualization of fiber surfaces without explicit extraction of fiber surfaces. Sakurai et al.~\cite{TA:Sakurai:2020:flexibleFiberSurfaces} extended the idea of flexible isosurfaces~\cite{TA:Carr:2010:flexibleIsosurfaces} to flexible fiber surfaces for the visualization of fiber surface components of interest without occlusion. Raith et al.~\cite{TA:Raith:2019:fiberSurfacesTensorFields} applied fiber surface extraction to tensor-field invariant spaces for the visualization of tensor fields. Tierny and Carr~\cite{TA:Tierny:2017:jacobiSetsReebSpaces} derived fiber surfaces around the Jacobi sets of bivariate fields (a conceptual equivalent of critical points of univariate fields) for computing fiber surface topological descriptors known as \textit{Reeb spaces}~\cite{TA:Edelsbrunner:2008:reebSpaces}. Our literature review covered a small subset of multivariate visualizations relevant to fiber surfaces. For a comprehensive overview of multivariate data analysis and visualization, we encourage readers to refer to the survey papers by Fuchs et al.~\cite{TA:Fuchs:2009:MultivariateSciVis} and Kehrer et al.~\cite{TA:Kehrer:2013:multifacetedDataVisSurvey}.

\subsection{Uncertainty Visualization}

To date, most of the research in uncertainty visualization has analyzed noise propagation in univariate data and the associated visualization algorithms. Specifically, these algorithms, including level~sets~\cite{TA:Rhodes:2003:uncertaintyVisIsosurfaceRendering, TA:Whitaker:2013:contourBoxPlots, TA:Kai:2013:nonparametricIsoVis, TA:Athawale:2016:nonparametricIsosurfaces, TA:Athawale2019ProbAsympDecider}, direct volume rendering~\cite{TA:Lundstrom:2007:probabilisticAnimationMedicalVolRendering, TA:Fout:2012:fuzzyVolumeRendering,TA:Shusen:2012:GMMdvr, TA:Sakhaee:2017:uncertainDVR, TA:Athawale:2021:nonparametricDVR}, and topology-based visualizations~\cite{TA:Wu:2013:ContourTreeUncertainty, TA:Gunther:2014:MandatoryCriticalPoints, TA:Favelier:2019:criticalPointVariabilityEnsembles, TA:Yan:2020:mergeTreeAverage, TA:Athawale:2022:uncertainMorseComplexes}, have been extensively studied in the context of uncertain univariate data. A few studies have investigated the uncertainty in visualizations of vector-field~\cite{TA:Lodha:1996:flowVisUncertainty, TA:Otto:2011:3dVectorFieldTopologyUncertainty, TA:2012:Schneider:ftleUncertainty, TA:Ferstl:2016:streamlineVariabilityVis, TA:Guo:2016:LyapunovUncertaintyUnsteadyFlow} and tensor-field~\cite{TA:Jones:2003:uncertaintyConeGlyphsTractography, TA:Jiao:2012:hardiDiffusionDataUncertaintyGlyph, TA:Siddiqui:2021:tensorImagingVisPipelineUncertainty} data.

There have been a few developments in uncertainty visualization of multivariate data. Zheng and Sadlo~\cite{TA:Zheng:2021:uncertainScatterPlots} proposed a framework for visualizing continuous scatterplots and uncertain fibers when the uncertainty in bivariate data is modeled with Gaussian distributions. Hazarika et al.~\cite{TA:Hazarika:2018:copulaBasedDistributionModels} characterized multivariate data distributions with copula-based models and proposed a sampling strategy for copula-based models to quantify the uncertainty of features such as level sets and vortices. Xie et al.~\cite{TA:Xie:2006:scatterplotPCPwithColorEncodedDataQuality} encoded data quality into scatterplots and parallel coordinate plots for multivariate data in the context of information visualization. Nagaraj et al.~\cite{TA:2011:Nagraj:gradientComparisonMultivariate} studied the consistency and inconsistency of gradient fields derived across attributes of multivariate data. In this paper, we take a step toward uncertainty quantification for multivariate data by
investigating uncertainty in fiber visualizations of bivariate data for parametric and nonparametric noise distributions and arbitrary FSCP shapes.

\section{Fiber Uncertainty for Independent Parametric Noise Models}\label{sec:parametric}

\fix{
In this section, we discuss two approaches, the closed-form integration and Monte Carlo integration, for fiber uncertainty computations. Although we present a closed-form fiber uncertainty quantification framework for integrable kernels (uniform, Epanechnikov, and Gaussian), a more generic Monte Carlo approach can be utilized to estimate fiber uncertainties for integrable as well as nonintegrable kernels.}

\subsection{Closed-Form Integration} \label{sec:closedFormIntegration}
\begin{figure}[!ht]
 \centering 
 \vspace{-2mm}
 \includegraphics[width=0.7\columnwidth]{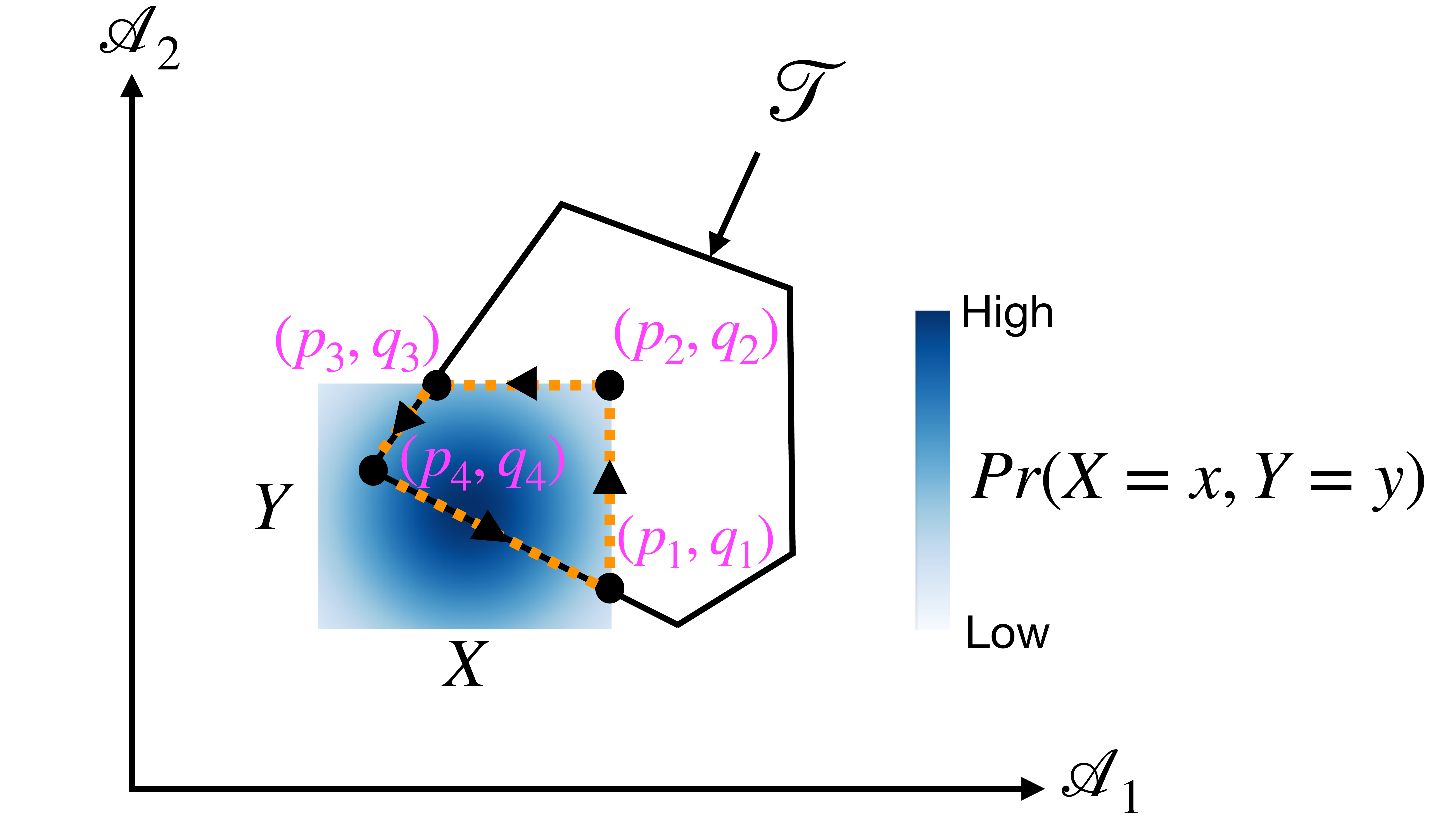}
  \vspace{-1mm}
 \caption{Illustration of interior probability computation for a single point of a spatial domain when FSCP has an arbitrary shape. $X$ and $Y$ denote the uncertain data ranges of a point, with blue denoting the joint probability distribution over the uncertain range. The polygon with black boundaries denotes trait $\mathcal{T}$. The intersection of the two regions is indicated by dotted orange lines, and the coordinates of the intersection are indicated by pairs ($p_i, q_i$). The probability of a point being in the interior of a fiber surface can be computed with Green's theorem by summing the line integrals of the probability density function (blue) along the edges of the intersection polygon in a counterclockwise direction (see arrow heads).}
 \label{fig:parametricTechnique}
\end{figure}

We address the challenge of uncertainty quantification of fibers, similar to Zheng and Sadlo~\cite{TA:Zheng:2021:uncertainScatterPlots}, for arbitrary shapes of FSCP. Specifically, we present a closed-form framework for computing interior probabilities (i.e., Pr($\mathcal{U}$ = $(X, Y)$ $\in \mathcal{T}$)) per grid vertex when data noise is \fix{modeled with the independent uniform, Epanechnikov, and Gaussian distributions, which are integrable}. The techniques detailed in this section are then leveraged as building blocks of our derivations for nonparametric noise models in \autoref{sec:nonparametricIndependent}. When the FSCP shape is not rectangular, the double integral in \autoref{eq:rectangularIntegration} cannot be utilized because it considers the ranges of two attributes. For FSCP with an arbitrary shape, we apply Green's theorem to integrate the joint probability distribution $\text{\fix{pdf}}_{X,Y}(x,y)$. Our approach is depicted in~\autoref{fig:parametricTechnique}.

In \autoref{fig:parametricTechnique}, the blue region indicates the joint probability distribution of random variables $X$ and $Y$ (i.e., $\text{\fix{pdf}}_{X,Y}(x,y)$). The polygon indicated by the orange dotted lines in \autoref{fig:parametricTechnique} results from the intersection of the uncertain region (blue) with trait $\mathcal{T}$. Let ($p_i, q_i$) denote the coordinates of the intersection polygon. We compute the interior probability (i.e., Pr($\mathcal{U}$ = $(X, Y)$ $\in \mathcal{T}$)) by integrating the joint probability distribution $\text{\fix{pdf}}_{X,Y} (x,y)$ over the intersection polygon (depicted by the orange dotted lines) using Green's theorem. 

To use Green's theorem, we compute the new polynomials, $L = (-1/2) \int \text{\fix{pdf}}_{X,Y} (x,y) dy$ and $M = (1/2) \int \text{\fix{pdf}}_{X,Y} (x,y) dx$.  Assuming that $X$ and $Y$ are independent, the $\text{\fix{pdf}}_{X,Y}(x,y)$ = $\text{\fix{pdf}}_{X}(x)\cdot \text{\fix{pdf}}_{Y}(y)$. Thus, the polynomials $L$ and $M$ can be rewritten as follows:
\begin{align*} \label{eq:polynomials}
L &= (-1/2) \int \text{\fix{pdf}}_{X}(x)\cdot \text{\fix{pdf}}_{Y}(y) dy \\
   & = (-1/2) \text{\fix{pdf}}_{X}(x) \big[ \int \text{\fix{pdf}}_{Y}(y) dy \big]\\
M & = (1/2) \int  \text{\fix{pdf}}_{X}(x)\cdot \text{\fix{pdf}}_{Y}(y) dx \\\
& = (1/2) \text{\fix{pdf}}_{Y}(y) \big[ \int  \text{\fix{pdf}}_{X}(x) dx \big].  \numberthis 
\end{align*}

The integral over a line segment with end coordinates ($p_1, q_1$) and ($p_2, q_2$), denoted by ${I}_{(p_1, q_1), (p_2, q_2)}$, can be computed as follows:

\begin{equation} \label{eq:lineIntegration}
{I}_{(p_1, q_1), (p_2, q_2)} =  \int_{p_1}^{p_2} L  dx + \int_{q_1}^{q_2} M dy.
\end{equation}
Substituting the expressions for $L$ and $M$ presented in \autoref{eq:polynomials} into \autoref{eq:lineIntegration}, we have the following:
\begin{align*} \label{eq:lineIntegrationFullyExpanded}
{I}_{(p_1, q_1), (p_2, q_2)} =  &\int_{x=p_1}^{x=p_2} (-1/2) \text{\fix{pdf}}_{X}(x) \big[ \int \text{\fix{pdf}}_{Y}(y) dy \big] dx +\\ 
&\int_{y=q_1}^{y=q_2} (1/2) \text{\fix{pdf}}_{Y}(y) \big[ \int  \text{\fix{pdf}}_{X}(x) dx \big] dy  \\
= & \frac{-1}{2} \int_{x=p_1}^{x=p_2} \text{\fix{pdf}}_{X}(x) dx \int \text{\fix{pdf}}_{Y} (y) dy  +\\ 
&  \frac{1}{2} \int_{y=q_1}^{y=q_2} \text{\fix{pdf}}_{Y}(y) dy \int \text{\fix{pdf}}_{X}(x) dx.  \numberthis 
\end{align*}
For each vertex ($p_{i}, q_{i}$) of a polygon, the line integral (\autoref{eq:lineIntegrationFullyExpanded}) is computed for the line segment that connects coordinates ($p_{i}, q_{i}$) and ($p_{(i\%N)+1}, q_{(i\%N)+1}$) in a counterclockwise direction (refer to the arrow heads of the polygon with dotted orange lines in \autoref{fig:parametricTechnique}), where $N$ indicates the number of vertices of the intersection polygon. The expressions of integration of the uniform, Epanechnikov, and Gaussian distribution functions (computed with Wolfram alpha~\cite{TA:Wolfram:2021:Mathematica}) are provided in \autoref{tab:integrationTable}. Finally, the interior probability, Pr($\mathcal{U}$ = ($X, Y)$ $\in \mathcal{T}$), can be computed using Green's theorem (i.e., by summing the line integrals [\autoref{eq:lineIntegrationFullyExpanded}] computed for each edge of the intersection polygon). \fix{Please refer to Sect. 1 of the supplementary material for a numerical integration experiment validating our derivation in \autoref{eq:lineIntegrationFullyExpanded}.}  
\fix{\subsection{Monte Carlo Integration}\label{sec:MonteCarloIntegration}
 In \autoref{sec:closedFormIntegration}, we derived interior probabilities in closed form for parametric distributions with the uniform, Epanechnikov, and Gaussian kernels. Such analytical derivations are possible since a closed-form integration exists for each of the three kernels (see \autoref{tab:integrationTable}). For parametric distributions that are difficult to integrate or do not have a closed-form integration, a more generic Monte Carlo integration can be employed to obtain an approximate solution. In the Monte Carlo approach for the independent noise assumption, $R$ samples are independently drawn from distributions $\text{\fix{pdf}}_X(x)$ and $\text{\fix{pdf}}_Y(y)$ at a vertex. If $S$ samples lie in the interior of trait $\mathcal{T}$, we quantify or estimate the interior probability as Pr($\mathcal{U}$ $\in \mathcal{T}$) = $S/R$. This estimation technique is similar to the probabilistic marching cubes~\cite{TA:Pothkow2011ProbMarchingCubes} technique, which estimates the spatial probability for level sets via Monte Carlo sampling of probability distributions. The accuracy of Monte Carlo solutions increases and converges with an increase in the number of samples.
}
\begin{table}[!t]
\begin{center}
\caption{\label{tab:integrationTable}\fix{Integration table}}
\begin{tabular}{| c | c | c |}
	
	\hline
	 Kernel & $K(x)$ & $\int K(x) dx$ \\ [5pt] \hline 
	&&\\[-1em]
	Uniform & $\frac{1}{2} \mathbbm{1}_{(|x| \le 1)}$    & $\frac{1}{2} x_{(|x| \le 1)}$  \\ [5pt]\hline
	&&\\[-1em] 
	Epanechnikov &  $\frac{3}{4} (1-x^2)_{(|x| \le 1)}$ & $\frac{3}{4} (x-\frac{x^3}{3})_{(|x| \le 1)}$\\ [5pt]\hline
	&&\\[-1em] 
	Gaussian & $\frac{1}{\sqrt{2\pi}} {e^{\frac{-1}{2} x^2}_{(\mu=0, \sigma=1)}}$ & $\frac{1}{2} erf (\frac{x}{\sqrt{2}})$ \\ [5pt]\hline	 
		
\end{tabular}
\end{center}
\end{table}

\section{Fiber Uncertainty for Nonparametric Noise Models}\label{sec:nonparametric}
\subsection{Independent Noise Assumption}\label{sec:nonparametricIndependent}

\fix{We will first discuss a closed-form integration approach for fiber uncertainty quantification.} For our closed-form approach, we utilize the \fix{analytical} derivations of independent parametric noise models proposed in \autoref{sec:closedFormIntegration} as building blocks for deriving polynomial integration of nonparametric noise models. In nonparametric density models, the probability distribution of each uncertain variable per vertex can be estimated by deriving a histogram or by using KDE~\cite{TA:Parzen:1968:ParzenWindow}. Let $\{x_{1}, x_{2}, .., x_{m}\}$ denote $m$ independently drawn samples from an unknown probability distribution $\text{\fix{pdf}}_{X}(x)$ of a random variable $X$. The $\text{\fix{pdf}}_X(x)$ can then be estimated from samples with KDE as follows:


\begin{equation} \label{eq:kde1}
\text{\fix{pdf}}_{X}(x) = \frac{1}{m} \sum_{i=1}^{i=m}K_{h_x}( x  -  x_{i}).
\end{equation}
The $h_x$ in \autoref{eq:kde1} denotes a non-negative bandwidth of a kernel $K$ associated with a sample. $K_{h_x}( x  -  x_{i})$ represents a scaled kernel, where $K_{h_x}( x  -  x_{i}) = \frac{1}{h_x} K(\frac{x  -  x_{i}}{h_x})$. The functions $K(x)$ in \autoref{tab:integrationTable} denote kernels with a unit bandwidth centered at $x_i=0$. We estimate the bandwidth of a kernel from samples using Silverman's rule of thumb~\cite{TA:1998:bandwidthSelection}. Similarly, for samples $\{y_{1}, y_{2}, .., y_{m}\}$ independently drawn from an unknown probability distribution $\text{\fix{pdf}}_{Y}(y)$, the probability distribution $\text{\fix{pdf}}_{Y}(y)$ for bandwidth $h_y$ can be estimated as follows:

\begin{equation} \label{eq:kde2}
\text{\fix{pdf}}_{Y}(y) = \frac{1}{m} \sum_{j=1}^{j=m}K_{h_y}( y  -  y_{j}).
\end{equation}
Substituting \autoref{eq:kde1} and \autoref{eq:kde2} in \autoref{eq:lineIntegrationFullyExpanded}, the line integral for the independent nonparametric density estimation can be written as
\begin{align*} \label{eq:lineIntegrationNonparametric}
&{I}_{(p_1, q_1), (p_2, q_2)} = \\ 
& \frac{-1}{2} \int_{x=p_1}^{x=p_2}  \frac{1}{m} \sum_{i=1}^{i=m}K_{h_x}( x  -  x_{i}) dx
 \int  \frac{1}{m} \sum_{j=1}^{j=m}K_{h_y}( y  -  y_{j})dy  +\\ 
&  \frac{1}{2} \int_{y=q_1}^{y=q_2}  \frac{1}{m} \sum_{j=1}^{j=m}K_{h_y}( y  -  y_{j})
 dy \int \frac{1}{m} \sum_{i=1}^{i=m}K_{h_x}( x  -  x_{i})dx.   \numberthis 
\end{align*}
Rearranging the order of summations in \autoref{eq:lineIntegrationNonparametric} gives us
\begin{align*} \label{eq:lineIntegrationNonparametricExpanded}
&{I}_{(p_1, q_1), (p_2, q_2)} = \\ 
& \frac{1}{m^2} \sum_{i=1}^{i=m} \sum_{j=1}^{j=m} \big[ \frac{-1}{2}  \int_{x=p_1}^{x=p_2}  K_{h_x}( x  -  x_{i}) dx
 \int  K_{h_y}( y  -  y_{j})dy  +\\ 
&  \frac{1}{2} \int_{y=q_1}^{y=q_2} K_{h_y}( y  -  y_{j})
 dy \int K_{h_x}( x  -  x_{i})dx \big].   \numberthis 
\end{align*}
\autoref{eq:lineIntegrationNonparametricExpanded} takes a form similar to the one for the parametric statistics presented in \autoref{eq:lineIntegrationFullyExpanded}, except that it loops through all pairs of kernels for random variables $X$ and $Y$. Kernel $K$ in \autoref{eq:lineIntegrationNonparametricExpanded} can be replaced with the uniform, Epanechnikov, or Gaussian functions presented in \autoref{tab:integrationTable}. 

Histograms could be considered as a special case of KDE with nonoverlapping uniform kernels associated with fixed-bin centers (precomputed from noise samples and user-specified number of bins) and with a bandwidth equivalent to the bin width. Thus, for histograms, looping through all pairs of bins of histograms of random variables $X$ and $Y$ is required for computing the line integral in \autoref{eq:lineIntegrationNonparametricExpanded}. Finally, the interior probability (i.e., Pr($\mathcal{U}$ = $(X, Y)$ $\in \mathcal{T}$)) can be computed using Green's theorem by summing line integrals for all edges of the intersection polygon, as illustrated in \autoref{fig:parametricTechnique}. 

\fix{When kernel $K$ of a nonparametric distribution is difficult to integrate or nonintegrable, a Monte Carlo approach similar to the one for independent parametric distributions (\autoref{sec:MonteCarloIntegration}) can be employed. In other words, for $R$ samples independently drawn from nonparametric distributions $\text{\fix{pdf}}_X(x)$ and $\text{\fix{pdf}}_Y(y)$ with the base kernel $K$, the interior probability can be estimated as Pr($\mathcal{U}$ $\in \mathcal{T}$) = $S/R$ if $S$ samples lie in the interior of trait $\mathcal{T}$.}

\subsection{Correlated Noise Assumption}\label{sec:nonparametricCorrelated}
When random variables $X$ and $Y$ are correlated, $\text{\fix{pdf}}_{X,Y} (x,y)$ is not equal to the product $\text{\fix{pdf}}_X(x) \cdot \text{\fix{pdf}}_Y(y)$. In the correlated noise case, 2D histograms or bivariate KDE can be used to capture the correlation between variables $X$ and $Y$ and the multimodality of their probability distributions. Bivariate KDE extends the idea of univariate KDE in \autoref{eq:kde1} to two variables~\cite{TA:1993:Wand:kdeBivariate}. \fix{An example of 2D histograms and bivariate KDE is provided in Sect. 2 of the supplementary material.} 

For a 2D histogram, the interior probability (i.e., Pr($\mathcal{U}$ = $(X, Y)$ $\in \mathcal{T}$)) can be computed in closed form. Specifically, we loop through each 2D bin of a 2D histogram and quantify the amount of bin overlap with trait $\mathcal{T}$ by taking into account the bin extent. We then scale the quantified overlap for a 2D bin with the weight of a bin computed from a 2D histogram. Our approach here is similar to the histogram approach for the independent nonparametric case (\autoref{sec:nonparametricIndependent}), except that the weight of each bin is not a product $\text{\fix{pdf}}_X(x) \cdot \text{\fix{pdf}}_Y(y)$, but the weight computed based on a 2D histogram.

In the case of a bivariate KDE, we propose a numerical integration approach similar to the one presented in \fix{Sect. 1 of the supplementary material}. Specifically, we discretize the uncertain ranges of random variables $X$ and $Y$ with a uniform grid. We then estimate the density at each grid vertex using a bivariate KDE. We use the \texttt{gaussian\_kde} function from the Python SciPy package~\cite{TA:Virtanen:2020:SciPy} for bivariate KDE. We then perform a point-in-polygon test for user-specified FSCP at each grid vertex and sum the weights of all vertices that lie inside FSCP or trait $\mathcal{T}$ to compute the interior probability (i.e., Pr($\mathcal{U}$ = $(X, Y)$ $\in \mathcal{T}$)).
\vspace{-1.2mm}
\section{Memory and Computational Complexity}\label{sec:complexity}

\begin{figure*}[!ht]
 \centering 
 \includegraphics[width=\textwidth]{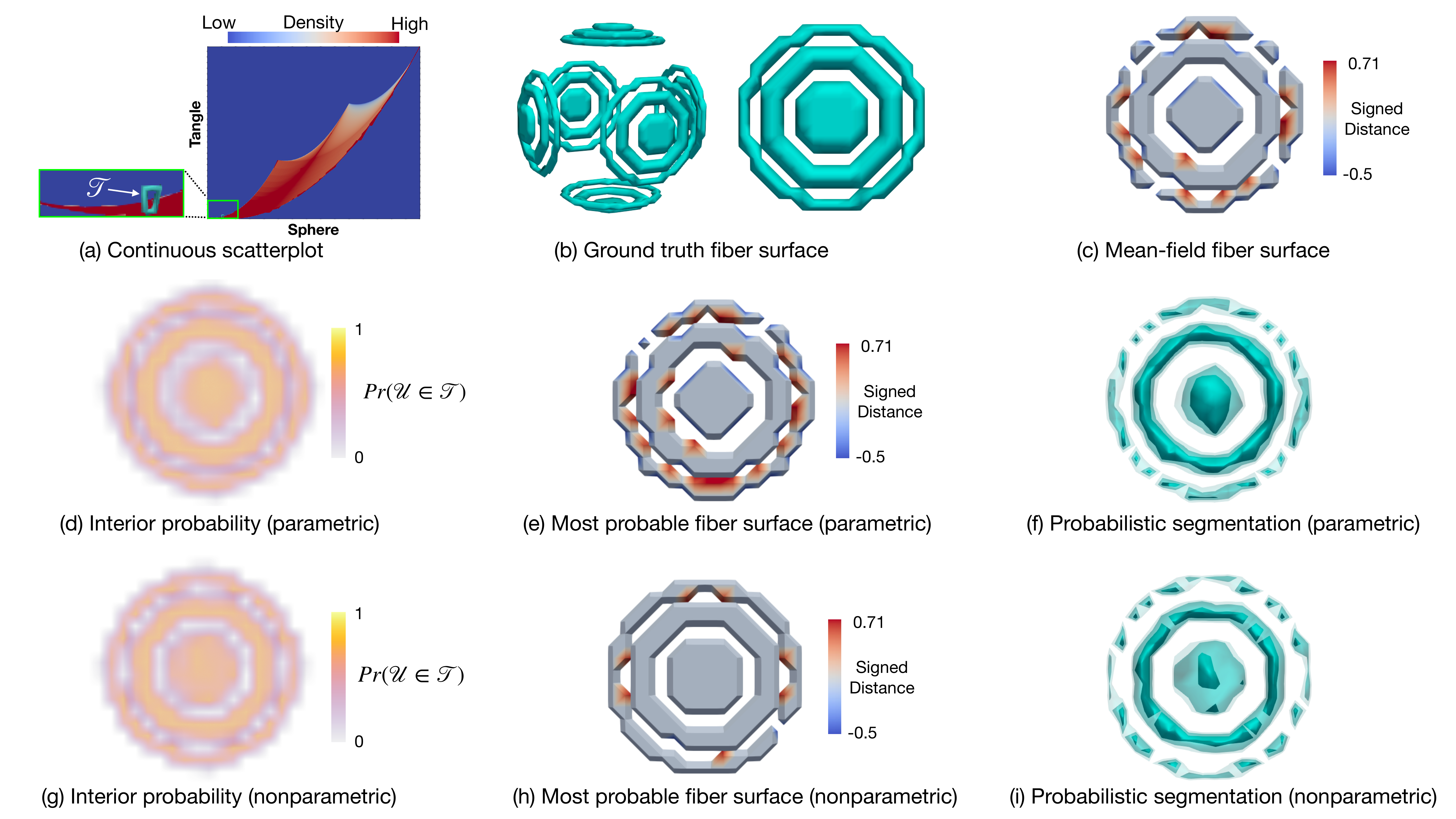}
  \vspace{-5mm}
 \caption{Fiber visualizations for the synthetic tangle-sphere dataset. A continuous scatterplot for the dataset is visualized in image (a) with trait $\mathcal{T}$ indicated by a cyan polygon. Image (b) visualizes the ground truth fiber surface that corresponds to trait $\mathcal{T}$. The ground truth dataset is mixed with random samples drawn from a bimodal noise distribution to generate an ensemble. Image (c) visualizes the mean-field fiber surface with a color-mapped signed distance from the ground truth fiber surface. The volume rendering of the interior probability volume, most probable fiber surface, and probabilistic segmentation are visualized for parametric (d)--(f) and nonparametric (g)--(i) statistical models. The volume rendering for nonparametric noise models visually represents the ground truth fiber positions more accurately than the volume rendering for parametric models (see \autoref{fig:errorAnalysisTangleSphereIntersection} for quantitative analysis). Similarly, the most probable fiber surface for nonparametric models is spatially closer to the ground truth fiber surface than the surfaces for the mean-field and parametric models, as is evident by the color-mapped signed distance. The probabilistic segmentation provides insight into the positions with relatively high (opaque) or low (translucent) probability of fiber existence.}
 \label{fig:tangleSphereIntersection}
\end{figure*}

Here, we discuss the memory and computational complexity of our proposed parametric and nonparametric statistical models. Let the size of uncertain input data be $D_1 \times D_2 \times D_3 \times V \times M$, where $D_i$ denotes the $i^{th}$ dimension of a uniform grid that represents domain $\mathcal{D}$, $V$ indicates the number of variables ($V=2$ for bivariate data), and $M$ is the number of ensemble members or noise samples that represent uncertainty in the data. For \fix{closed-form derivations with the independent} parametric noise models in \fix{\autoref{sec:closedFormIntegration}}, we summarize $M$ noise samples with two parameters per grid vertex. For example, we store the mean and width per vertex for a uniform distribution and the mean and standard deviation for a Gaussian distribution. Thus, for parametric noise models, the memory consumed is $D_1 \times D_2 \times D_3 \times (V=2) \times 2$, which amounts to data reduction by a factor of $M/2$. To compute the interior probability volume for parametric noise models, one polynomial integration is performed per trait $\mathcal{T} \subset \mathcal{A}$ per grid vertex if uncertain data at a vertex intersect trait $\mathcal{T}$. Thus, the worst-case computational complexity is proportional to the grid size (i.e., $D_1 \times D_2 \times D_3$).

For \fix{closed-form derivations with} the independent nonparametric statistical approach (\autoref{sec:nonparametricIndependent}), the probability density per vertex can be estimated with KDE or histograms. For KDE, if a kernel is associated with each of the $M$ noise samples per vertex and per variable, then the memory requirement is the same as the original data (i.e., $D_1 \times D_2 \times D_3 \times (V=2) \times M$). If nonparametric density is estimated with a histogram with $B$ bins per vertex and per variable, then memory consumption is reduced to $D_1 \times D_2 \times D_3 \times (V=2) \times B$, which amounts to data reduction by a factor of $M/B$. To compute the interior probability volume, the number of computations per vertex increases quadratically with the number of kernels or histogram bins owing to the nested summation in \autoref{eq:lineIntegrationNonparametricExpanded} (see the experiment in  \autoref{fig:accuracyTiming}).

\fix{For the more generic Monte Carlo approach described in \autoref{sec:MonteCarloIntegration} and \autoref{sec:nonparametricIndependent}, the number of computations grows in proportion to the grid size and number of samples drawn per grid vertex (i.e., $D_1 \times D_2 \times D_3 \times S$).}
In the case of the correlated nonparametric statistical approach (\autoref{sec:nonparametricCorrelated}), the space complexity is again equivalent to that of the independent nonparametric approach. For 2D histograms, the number of computations grows quadratically with the number of bins because we loop through each 2D bin. For bivariate KDE, we perform numerical integration by discretizing the uncertain data range at a vertex with a uniform grid of resolution, $i \times i$. Thus, the worst-case time complexity is proportional to $D_1 \times D_2 \times D_3 \times i \times i$ per trait $\mathcal{T}$ if the uncertain data range at each grid vertex intersects trait $\mathcal{T}$.

\section{Visualization of Uncertain Fibers}\label{sec:visualizationTechniques}
We visualize \fix{the positional uncertainty of fibers, i.e., the interior probability volumes derived using our proposed parametric (\autoref{sec:parametric}) and nonparametric (\autoref{sec:nonparametric}) statistics,} via most probable fiber surface extraction, direct volume rendering, and probabilistic segmentation. Athawale et al. proposed a vertex-based classification framework~\cite{TA:Athawale:2016:nonparametricIsosurfaces} for extracting the most probable isosurface from uncertain scalar fields. \fix{We extend their idea for extracting the most probable fiber surface from uncertain bivariate data.} Specifically, we derive the interior probability Pr($\mathcal{U}$ = $(X, Y)$ $\in \mathcal{T}$) per vertex, as described in \autoref{sec:parametric} and \autoref{sec:nonparametric}. We then probabilistically predict the vertex sign as $+$ if Pr($\mathcal{U}$ $\in \mathcal{T}) \ge 0.5$; otherwise, we predict the vertex sign to be $-$. The isosurface extracted from a grid with this sign classification indicates the most probable fiber surface topology.

We visualize the positional uncertainty of fibers via direct volume rendering of interior probability volumes derived using parametric and nonparametric statistics, similar to the visualizations by Zheng and Sadlo~\cite{TA:Zheng:2021:uncertainScatterPlots}. We render the positions with relatively high interior probability using higher opacity and the ones with relatively low  interior probability using lower opacity. We visualize the probabilistic segmentation by thresholding computed interior probabilities. Specifically, for threshold $t$, we visualize the positions with Pr($\mathcal{U}$ $\in \mathcal{T}) >= t$. Such visualizations provide probabilistic insight into fiber positions.

\section{Results and Discussions}\label{sec:results}

\begin{figure}[!hb]
 \includegraphics[width=1\columnwidth]{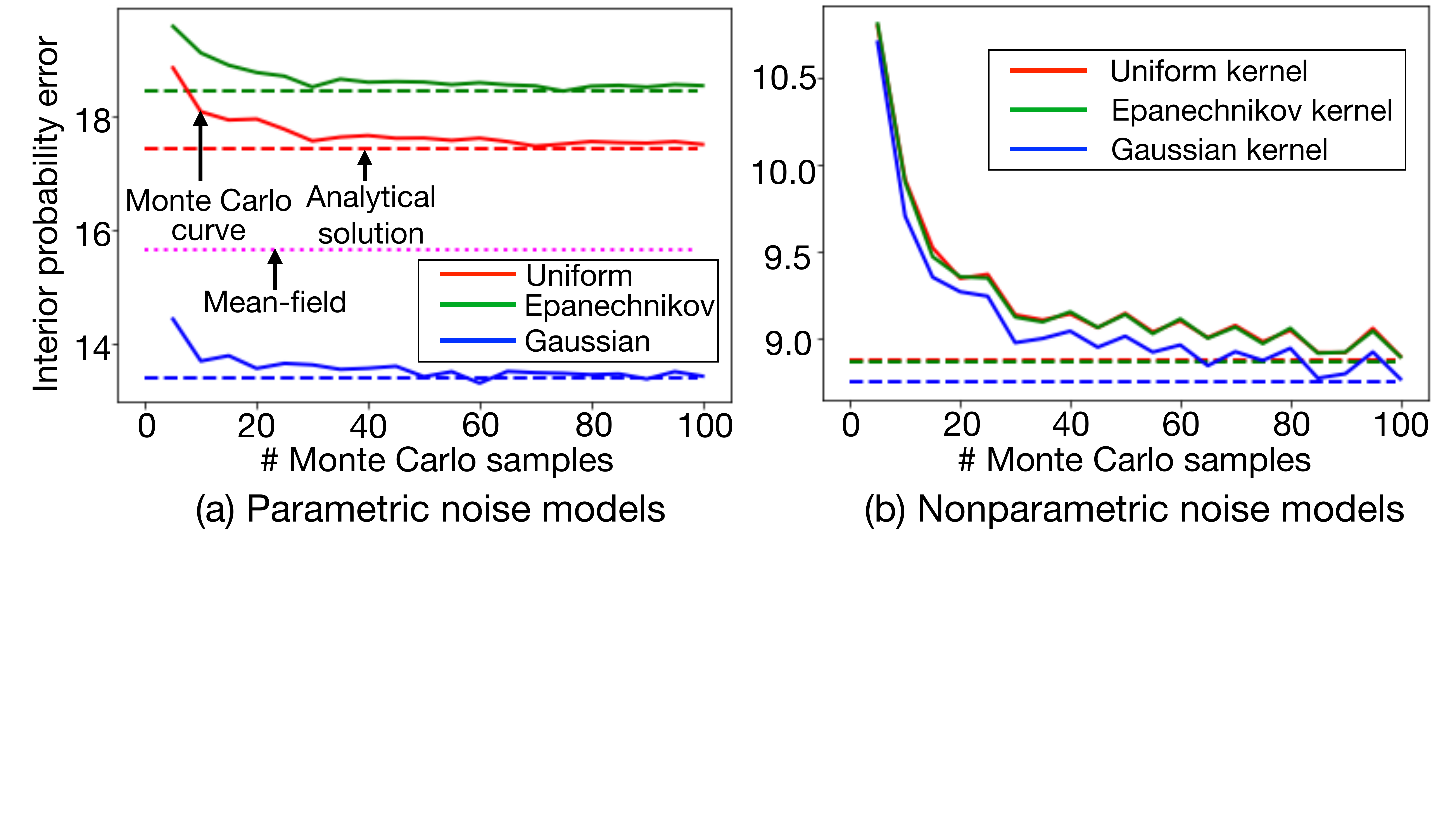}
  \vspace{-22mm}
 \caption{Quantitative error analysis of interior probability computations for independent parametric and nonparametric noise models with respect to the ground truth interior probabilities. The solid lines plot the error as a function of the number of Monte Carlo samples. The magenta dotted line denotes the error for the mean-field, and the dashed lines denote the error for our proposed analytical solutions.}
  \label{fig:errorAnalysisTangleSphereIntersection}
\end{figure}

We validate our techniques and present their effectiveness via experiments on synthetic and simulation datasets. 

\subsection{Synthetic Data}\label{sec:synthetic}

\subsubsection{Independent Noise Models}
We present the results of our proposed closed-form independent parametric (\autoref{sec:parametric}) and nonparametric (\autoref{sec:nonparametricIndependent}) statistical models for the synthetic data in \autoref{fig:tangleSphereIntersection}. For our experiment, we sample the synthetic sphere and tangle~\cite{TA:Knoll:2009:RaytracingImplicitSurfaces} functions on a 3D grid with resolution $64\times64\times64$. \autoref{fig:tangleSphereIntersection}a visualizes a 2D \fix{continuous} scatterplot of sphere and tangle values plotted on the horizontal and vertical axes, respectively, for each grid 
vertex. The \fix{cyan} polygon in \autoref{fig:tangleSphereIntersection}a denotes FSCP or trait $\mathcal{T}$, and the corresponding fiber surface is visualized in \autoref{fig:tangleSphereIntersection}b, which we treat as the reference. 

We mix the synthetic data sampled on a grid with noise to generate an ensemble of $40$ members. Specifically, we draw samples from a bimodal probability distribution, in which the mode with 80\% cumulative probability density is centered around the ground truth, and the mode with 20\% cumulative probability density is situated relatively far away from the ground truth. Thus, the noise samples from the mode with 20\% cumulative probability density denote the outlier samples. We study the uncertainty in fiber positions arising from the ensemble.

\autoref{fig:tangleSphereIntersection}c visualizes the fiber surface for the mean-field. The mean-field fiber surface is color-mapped with respect to its signed distance~\cite{TA:2005:signedDistance} from the ground truth fiber surface. Figs.~\ref{fig:tangleSphereIntersection}d--\ref{fig:tangleSphereIntersection}f visualize the results for the independent Gaussian noise assumption. \autoref{fig:tangleSphereIntersection}d visualizes the interior probability volume (i.e., Pr($\mathcal{U}$ $\in \mathcal{T}$)) per grid vertex via direct volume rendering, in which yellow with higher opacity denotes positions with relatively high interior probabilities, and white with lower opacity denotes positions with relatively low interior probabilities. \autoref{fig:tangleSphereIntersection}e visualizes the most probable fiber surface computed using the vertex-based classification (\autoref{sec:visualizationTechniques}) color-mapped with the signed distance from the ground truth fiber surface. \autoref{fig:tangleSphereIntersection}f visualizes the probabilistic segmentation (\autoref{sec:visualizationTechniques}) extracted for isovalues $0.9$, $0.75$, and $0.6$, and mapped to high, moderate, and low opacity, respectively. The probabilistic segmentation provides insight into the probability of fiber positions being in the interior of a fiber surface for different segments. Figs.~\ref{fig:tangleSphereIntersection}g--\ref{fig:tangleSphereIntersection}i visualize results similar to Figs.~\ref{fig:tangleSphereIntersection}d--\ref{fig:tangleSphereIntersection}f for the independent nonparametric KDE with the Gaussian base kernel. 

The most probable fiber surface for the nonparametric noise assumption (\autoref{fig:tangleSphereIntersection}h) is spatially closer to the ground truth fiber surface (\autoref{fig:tangleSphereIntersection}b) when compared with the mean-field fiber surface (\autoref{fig:tangleSphereIntersection}c) and the most probable fiber surface for the parametric assumption (\autoref{fig:tangleSphereIntersection}e), as is evident by the color-mapped signed distance. The result implies a greater accuracy of interior
probability computations per grid vertex with the nonparametric approach than with the mean-field or parametric approach. This result is expected because the nonparametric models can capture the bimodality of the underlying noise distribution more accurately than the parametric models, which results in nonparametric models that are less sensitive to outliers than the parametric models.

We further confirm the enhanced accuracy of nonparametric models compared to mean-field and parametric models using a quantitative error analysis, as shown in \autoref{fig:errorAnalysisTangleSphereIntersection}. Specifically, we compute and plot a Euclidean 2-norm of the difference between the interior probability volume (i.e., Pr($\mathcal{U}$ $\in \mathcal{T}$) per vertex)
computed using different statistical techniques and the interior probability volume for the ground truth data, which we refer to as {\em interior probability error}. In the case of the ground truth data, the interior probability is $1$ when data at a vertex lie inside trait $\mathcal{T}$; otherwise, it is $0$.~\autoref{fig:errorAnalysisTangleSphereIntersection}a
visualizes the results for parametric noise models with uniform (red), Epanechnikov (green), and Gaussian (blue) assumptions.~\autoref{fig:errorAnalysisTangleSphereIntersection}b
visualizes the results for nonparametric KDE with uniform~(red), Epanechnikov~(green), and Gaussian~(blue) base kernels. The dotted magenta line in \autoref{fig:errorAnalysisTangleSphereIntersection}a represents the error for the mean-field. In \autoref{fig:errorAnalysisTangleSphereIntersection}, we observe that the error computed for the nonparametric approach (\autoref{fig:errorAnalysisTangleSphereIntersection}b) is consistently lower than the error computed for the parametric density assumption and mean-field (\autoref{fig:errorAnalysisTangleSphereIntersection}a). The Gaussian kernel for the parametric and nonparametric statistics has the highest computational accuracy when compared with the uniform and Epanechnikov kernels.

We compare the results of the Monte Carlo sampling technique \fix{(\autoref{sec:MonteCarloIntegration} and \autoref{sec:nonparametricIndependent})} with our closed-form solutions in \autoref{fig:errorAnalysisTangleSphereIntersection}. In \autoref{fig:errorAnalysisTangleSphereIntersection}, the dashed lines denote the fixed error corresponding to our proposed closed-form solutions, and the solid curves plot the error as a function of the number of Monte Carlo samples. The error computed for the Monte Carlo solutions (solid curves) converges to the one computed for our proposed analytical solutions (dashed lines) as we increase the number of Monte Carlo samples. Such convergence confirms the correctness of our derivations (Equations~\ref{eq:lineIntegrationFullyExpanded}~and~\ref{eq:lineIntegrationNonparametricExpanded}) and implementations.


\begin{figure}[!t]
 \centering 
 \vspace{-3.5mm}
 \includegraphics[width=\columnwidth]{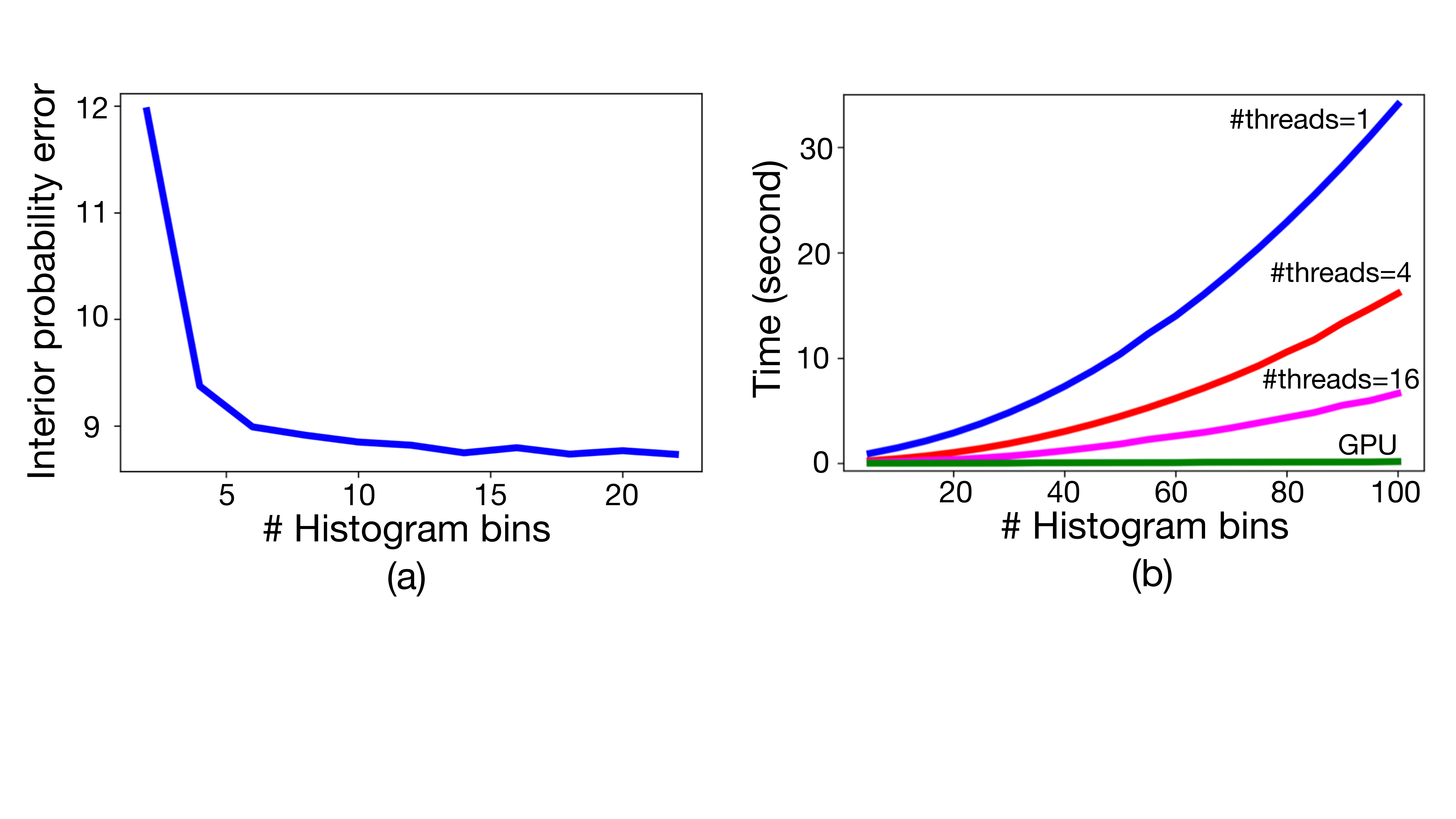}
 \vspace{-18mm}
 \caption{The (a) accuracy vs. (b) computational cost curves for interior probability computation using the nonparametric histogram model. The computational cost grows quadratically, but the accuracy improves as the number of histogram bins increases. \fix{The parallel implementation with multiple openMP threads (red and magenta curves) and GPU (green curve) provides computational speed-up.}}
  \label{fig:accuracyTiming}
\end{figure}

~\autoref{fig:accuracyTiming} visualizes the accuracy and time curves as a result of histogram rebinning for the independent nonparametric noise assumption. At each grid vertex of the domain, we perform nonparametric density estimation by computing a histogram of ensemble members followed by the computation of interior probability. We then compute and plot the interior probability error (i.e., the Euclidean 2-norm of the difference between the interior probability volumes for the histogram noise model and the ground truth data). The number of bins determines the memory and computational requirements of the proposed nonparametric approach (\autoref{sec:complexity}) and impacts the accuracy. As illustrated in \autoref{fig:accuracyTiming}a, the error reduces as the number of histogram bins increases. Note the sharp decline in errors for a relatively small number of bins. On the other hand, the computational complexity appears to grow quadratically, as we had anticipated. Since the interior probability at each grid vertex can be computed independently, the computations can be parallelized.  As illustrated in \autoref{fig:accuracyTiming}b, the red and magenta curves indicate the speed-up achieved with the OpenMP C++ parallel implementation with four and $16$ threads, respectively. \fix{The GPU CUDA version of our code achieved 23X average speed-up with the NVIDIA V100 graphics card (the green curve in \autoref{fig:accuracyTiming}b) compared to the Power9 CPU with $16$ openMP threads. The computing resources are courtesy of the Summit Supercomputer at the Oak Ridge National Laboratory.}

\subsubsection{Correlated Noise Models}
\begin{figure}[!b]
 \centering 
 \vspace{-2mm}
 \includegraphics[width=\columnwidth]{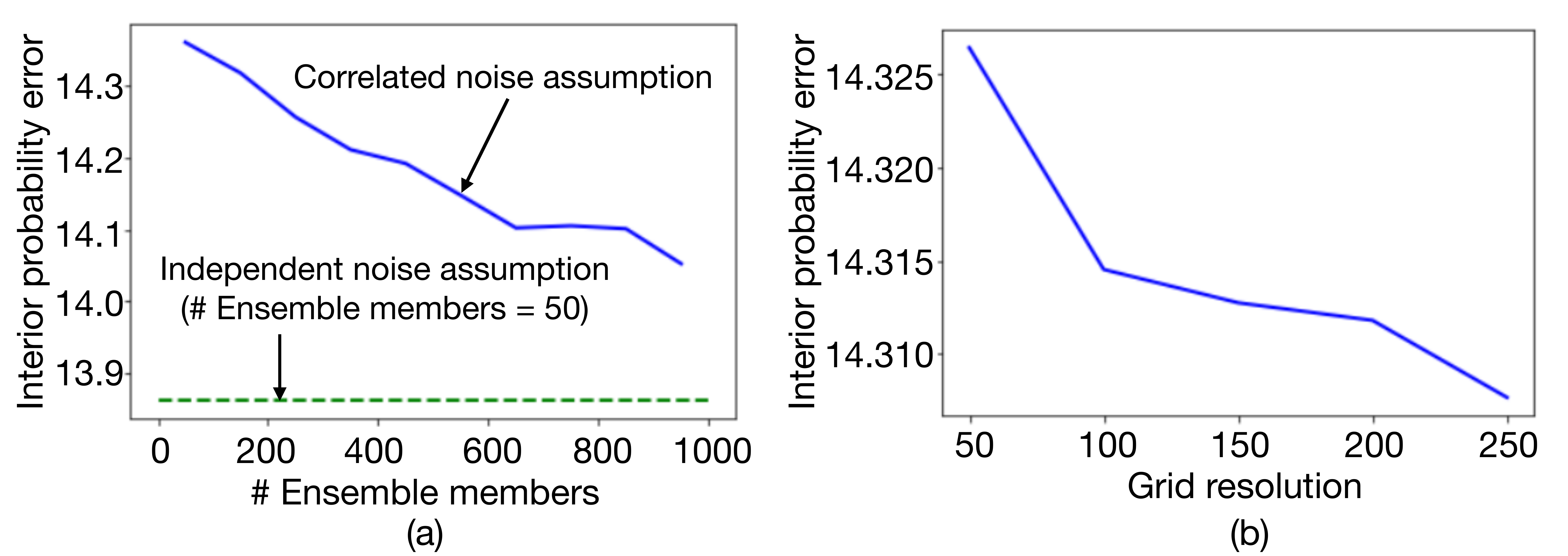}
 \vspace{-4mm}
 \caption{Interior probability error (blue curves) for the correlated noise assumption plotted as a function of the (a) number of ensemble members and (b) grid resolution for numerical integration.}
  \label{fig:correlatedErrorRate}
\end{figure}

We study the correlated noise models (\autoref{sec:nonparametricCorrelated}) via a synthetic experiment. The experimental settings are similar to the ones for the independent noise, except the bimodal noise now comprises two {\em correlated} Gaussians \fix{(e.g., Fig. 2a in the supplementary material)} with the correlation specified by covariance matrices. A correlated Gaussian with an 80\% probability concentration is situated close to the ground truth, and one with 20\% probability concentration is situated relatively far away from the ground truth (denoting the outliers). The noise samples drawn from such bimodal distributions are mixed with the ground truth to generate an ensemble.

We observe that the ability of bivariate KDE or 2D histograms to reliably capture the correlation between random variables $X$ and $Y$ at each grid vertex is influenced by the number of ensemble members and by the grid resolution used for numerical integration of bivariate KDE. When the number of noise samples is relatively large (e.g., $10,000$ samples for bivariate KDE estimation \fix{in Fig. 2c of the supplementary material}), the correlation is captured reliably. For fewer samples, the correlation might not be captured well and might lead to relatively large errors. As shown in \autoref{fig:correlatedErrorRate}a, the interior probability error (blue curve) reduces with an increase in the number of ensemble members. The interior probabilities computed with the independent nonparametric noise assumption (the green dotted line in \autoref{fig:correlatedErrorRate}a) appear to be more accurate than the correlated noise assumption when the number of ensemble members is $50$.

The error for nonparametric models with the correlated or independent noise assumptions (\autoref{fig:correlatedErrorRate}a) is again smaller than the errors for the mean-field and parametric noise models owing to the higher robustness of nonparametric models to outliers. The mean-field for $50$ ensemble members resulted in an interior probability error equal to $16.79$. The parametric Gaussian noise assumption resulted in an interior probability error equal to $15.41$. The error for the mean-field and parametric noise assumption does not fluctuate much with an increase in the number of ensemble members.~\autoref{fig:correlatedErrorRate}b visualizes the error as a function of a grid resolution used for numerical integration in the case of bivariate KDE. The error reduces slowly with an increase in grid resolution.
\begin{figure}[!b]
\centering     
\vspace{-3mm}
\subfigure[\small \fix{Continuous scatterplot with nonrectangular trait $\mathcal{T}_1$ and rectangular trait $\mathcal{T}_2$}]{\label{fig:rectangleVsArbitraryCSP}\hspace{-2mm}\includegraphics[width=0.9\linewidth]{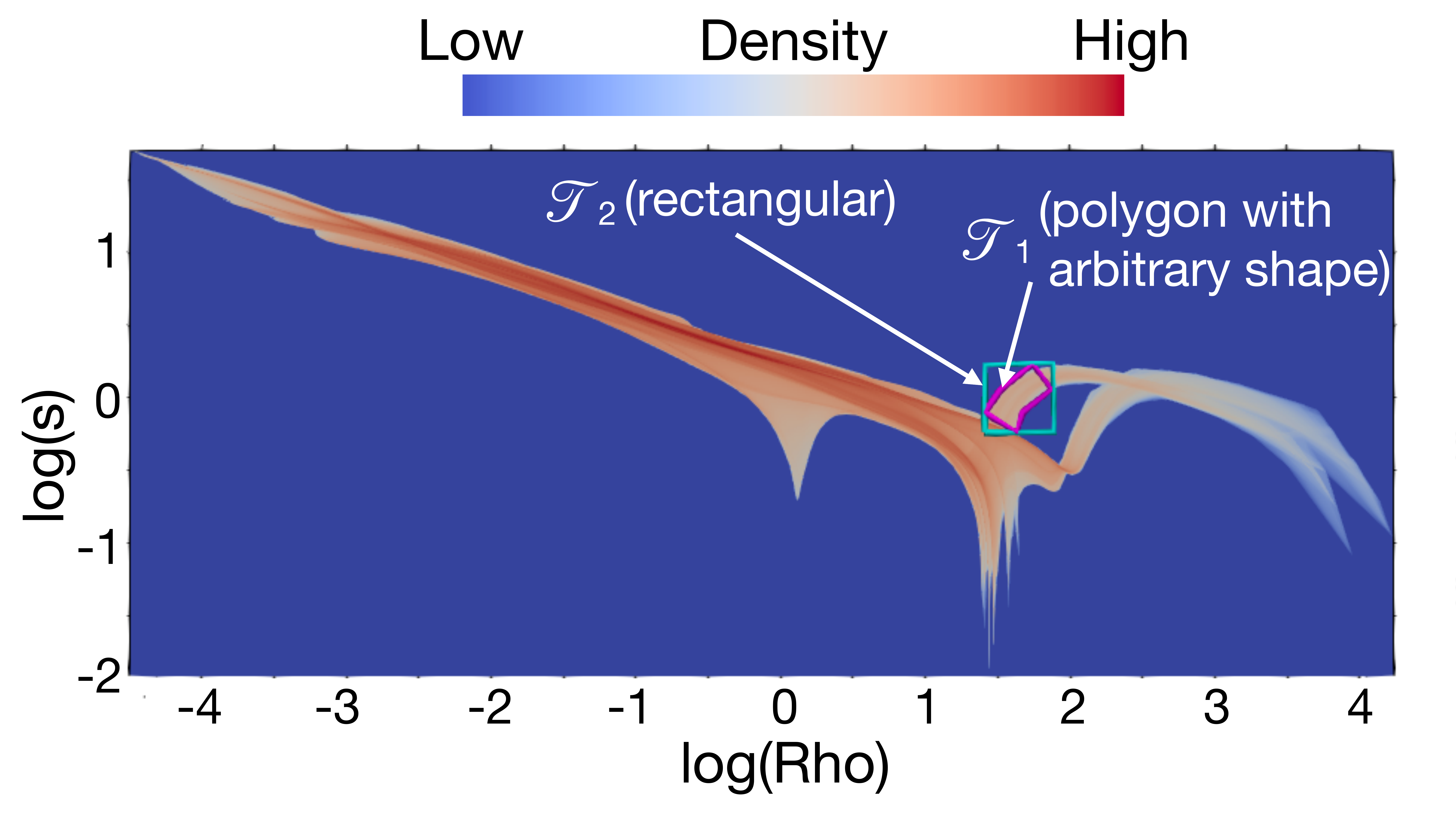}}
\subfigure[\small \fix{Fiber surface in the original data}]{\label{fig:rectangleVsArbitraryOriginal}\includegraphics[width=0.8\linewidth]{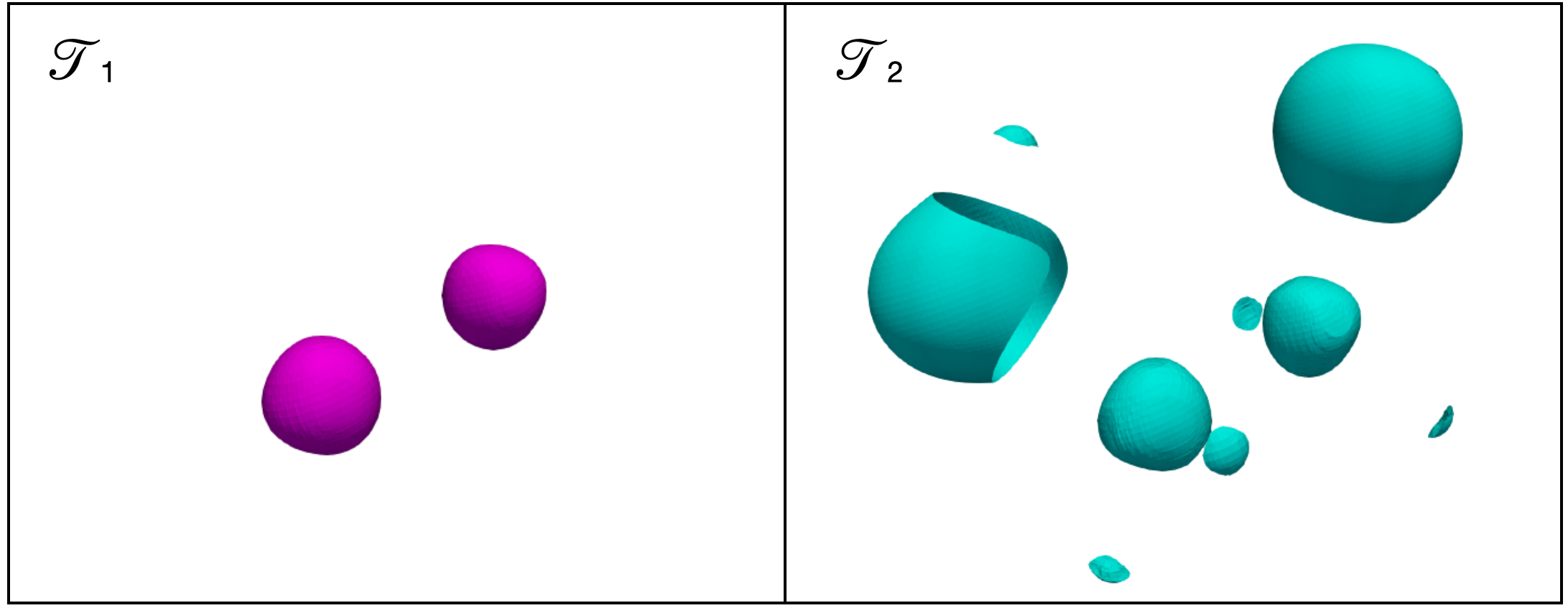}}
\subfigure[\small \fix{Fiber probabilities in the Gaussian-distributed hixel representation}]{\label{fig:rectangleVsArbitrarHixel}\includegraphics[width=0.8\linewidth]{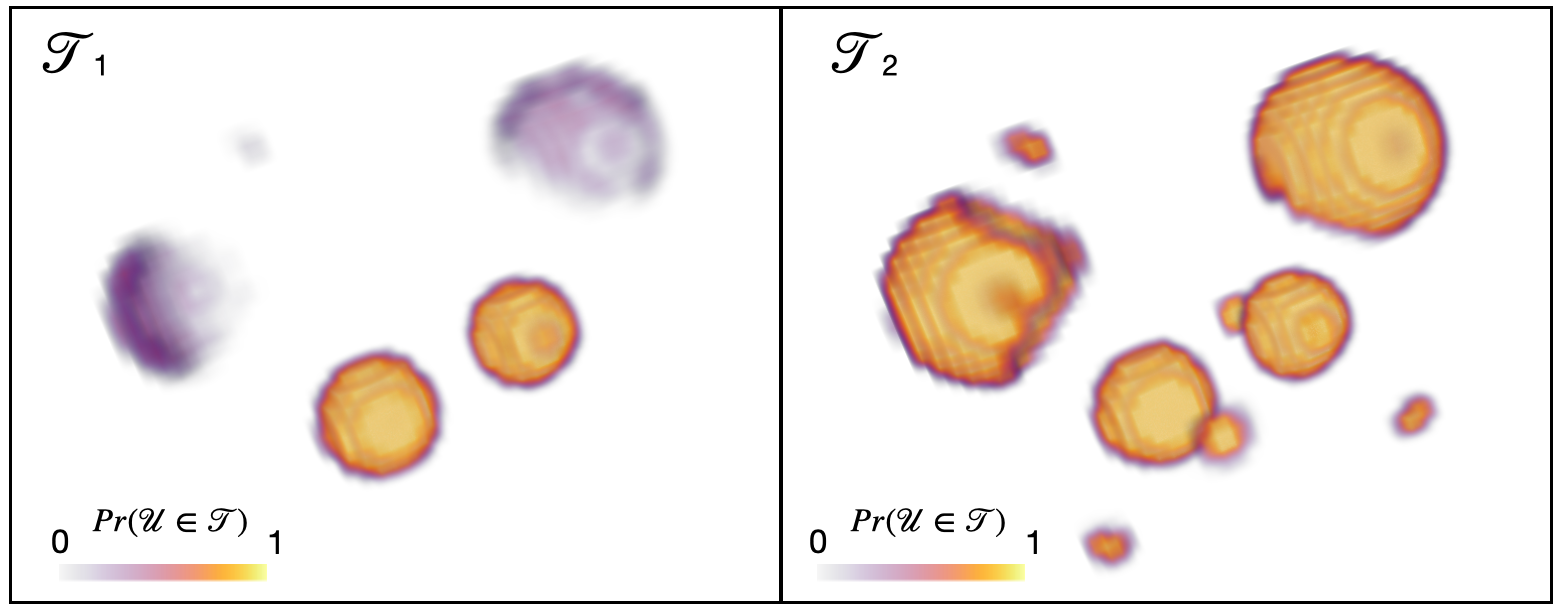}}
\caption{\fix{Uncertainty visualization of the ethanediol dataset with a rectangular~\cite{TA:Zheng:2021:uncertainScatterPlots} vs. arbitrary (our contribution) shape of the polygonal trait. The arbitrarily shaped trait $\mathcal{T}_1$ yields a more accurate extraction of carbon atom positions compared to the rectangular trait $\mathcal{T}_2$ for both (b) the original data and (c) uncertain hixel data}.}
\label{fig:rectangularVsArbitraryTraitEthaneDiol}
\end{figure}

\begin{figure*}[!ht]
 \centering 
 \includegraphics[width=0.9\textwidth]{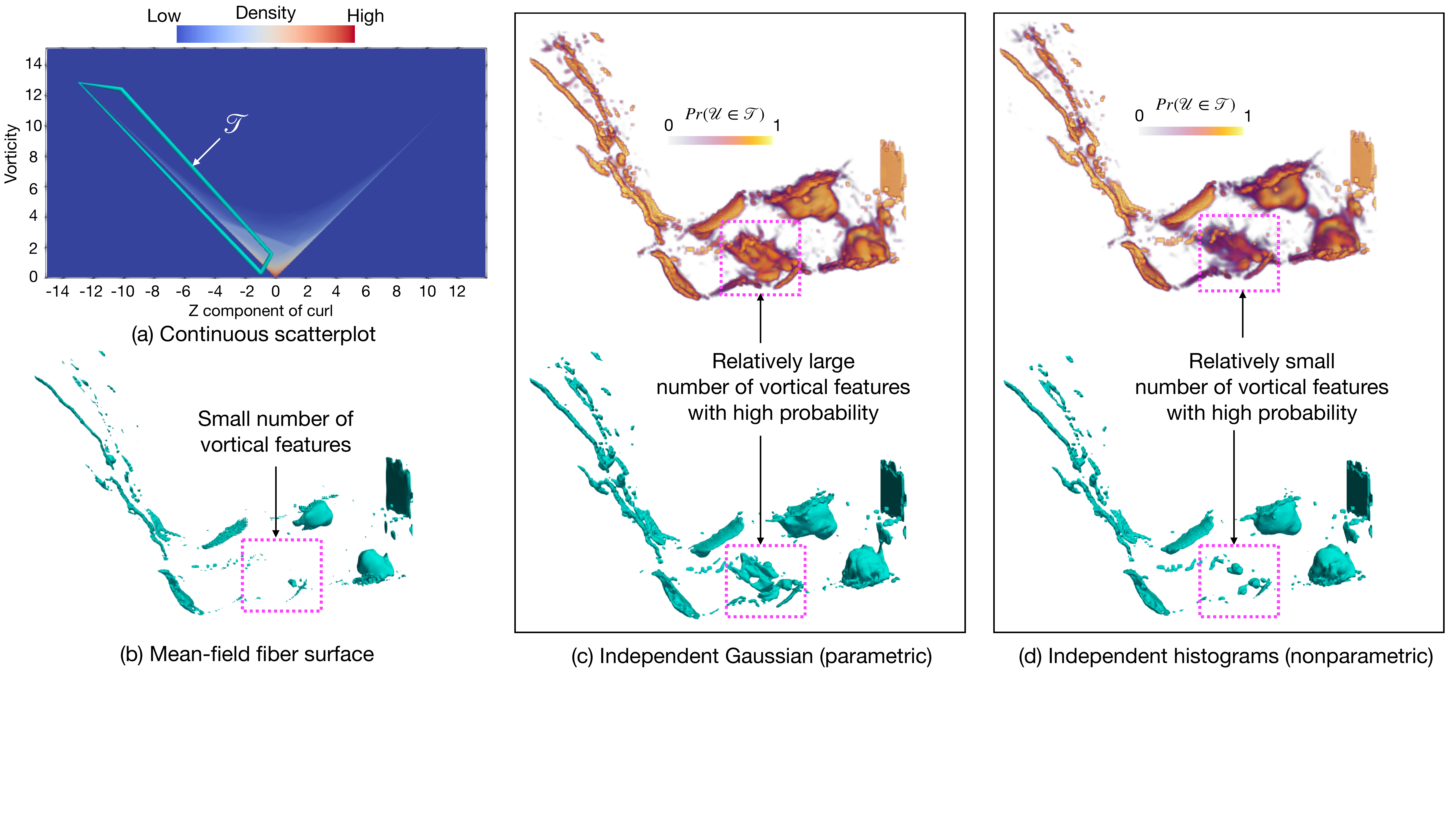}
 \vspace{-16mm}
 \caption{\fix{Parametric vs. nonparametric noise models for uncertainty visualization of vortical features of the Red Sea ensemble dataset over the Gulf of Aden. Fiber positions are visualized for trait $\mathcal{T}$ corresponding to anticyclonic (negative Z component of the curl of the velocity field) vortical features, as indicated by the cyan polygon in the continuous scatterplot shown in image (a). Image (b) visualizes the mean-field fiber surface. Images (c) and (d) visualize the results for the independent parametric (Gaussian) and independent nonparametric (histogram) noise models, respectively, with volume rendering of {\em interior probabilities} (top) and {\em a probabilistic segmentation} for isovalue $0.7$ (bottom). Although the three statistical models in images (b)-(d) exhibit overall consistency in the positions of eddy features, the uncertainty of eddy features seems prominent in the region enclosed by the magenta box. For instance, the parametric noise model yields multiple high-probability vortex features inside the magenta box, whereas the nonparametric noise model yields a relatively low probability of vortex features inside the magenta box}.}
 \label{fig:vorticityVisRedSea}
\end{figure*}
\vspace{-3.2mm}
\fix{\subsection{Rectangular Vs. Arbitrary Shape of Polygonal Traits}\label{sec:rectangularVsArbitrary}
The uncertainty visualization framework proposed by Zheng and Sadlo~\cite{TA:Zheng:2021:uncertainScatterPlots} is limited to a rectangular trait (\autoref{sec:stateOftheArtUncertaintyVisFibers}). Our proposed framework extends their work to a polygonal trait with arbitrary shape. ~\autoref{fig:rectangularVsArbitraryTraitEthaneDiol} demonstrates the advantage of an arbitrary polygon over rectangle for trait selection in the contexts of the original as well as uncertain data.~\autoref{fig:rectangularVsArbitraryTraitEthaneDiol}a visualizes a continuous scatterplot of the electron density (Rho) and reduced gradient (s) attributes of the ethanediol molecule. The main separating axis or diagonal of the continuous scatterplot (with strong red hue in~\autoref{fig:rectangularVsArbitraryTraitEthaneDiol}a) indicates regions with no chemical interactions. Thus, regions away from this axis are important for chemists in analyzing chemical interactions. The two traits $\mathcal{T}_1$ and $\mathcal{T}_2$ are selected in the continuous scatterplot space to enclose a nondiagonal feature. As observed in~\autoref{fig:rectangularVsArbitraryTraitEthaneDiol}a, the arbitrary shape of a polygon allows us to select a trait ($\mathcal{T}_1$ in magenta) that is closer to a nondiagonal feature than the rectangular shape of a trait ($\mathcal{T}_2$ in cyan).} 

\fix{The fiber surfaces corresponding to traits $\mathcal{T}_1$ and $\mathcal{T}_2$ are visualized for the original data in~\autoref{fig:rectangularVsArbitraryTraitEthaneDiol}b. The fiber surface for the nonrectangular trait $\mathcal{T}_1$ (the left image in~\autoref{fig:rectangularVsArbitraryTraitEthaneDiol}b) segments out carbon atoms more accurately than the fiber surface for the rectangular trait $\mathcal{T}_2$ (the right image in~\autoref{fig:rectangularVsArbitraryTraitEthaneDiol}b). Next, we reduce the original data with the Gaussian-distributed hixel representation~\cite{TA:Thompson:2011:hixelVis}. Specifically, we partition the original data into blocks of size $2 \times 2 \times 2$ and represent each block as a Gaussian distribution with mean and standard deviation. The fiber surfaces corresponding to traits $\mathcal{T}_1$ and $\mathcal{T}_2$ are visualized for the hixel representation in~\autoref{fig:rectangularVsArbitraryTraitEthaneDiol}c. The carbon atoms are again segmented well for trait $\mathcal{T}_1$ with the high probability (yellow) regions (the left image of~\autoref{fig:rectangularVsArbitraryTraitEthaneDiol}c). The two newly formed blobs in the left image of ~\autoref{fig:rectangularVsArbitraryTraitEthaneDiol}c that do not exist in the left image of ~\autoref{fig:rectangularVsArbitraryTraitEthaneDiol}b are due to the uncertainty in data arising from the hixel representation, but they have a very low probability (purple) of existence. As observed in the right image of~\autoref{fig:rectangularVsArbitraryTraitEthaneDiol}c, the rectangular trait $\mathcal{T}_2$ results in an overestimation of features, i.e., a multiple high probability (yellow) regions, that do not quite correctly represent carbon atoms.} 


\begin{figure*}[!ht]
 \centering 
 \includegraphics[width=\textwidth]{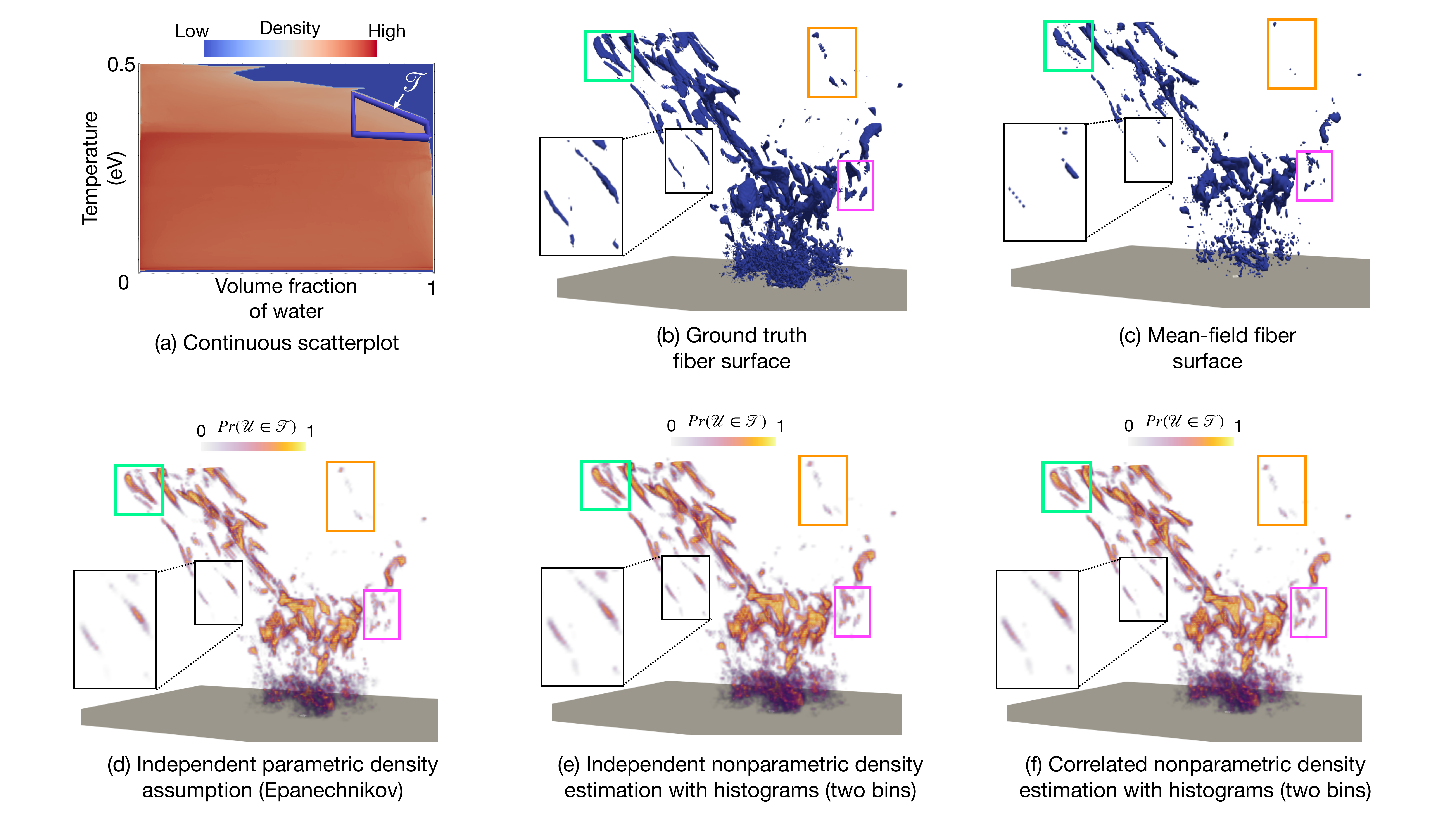}
 \vspace{-5mm}
 \caption{Fiber visualizations for the deep-water asteroid impact simulation dataset comprising the temperature and volume fraction of water variables. Image (a) visualizes a continuous scatterplot of the two variables with trait $\mathcal{T}$, as indicated by a dark blue polygon. Image (b) visualizes the ground truth fiber surface that corresponds to trait $\mathcal{T}$. The data are reduced with the hixel representation, where we partition the data into blocks of size $2\times2\times2$, and each block is summarized with a probability distribution. The fiber positions are visualized for the hixel data using different statistical techniques in (c)--(f) for the trait $\mathcal{T}$. Both parametric and nonparametric noise models show enhanced feature recovery when compared to the mean-field, as illustrated by the colored boxes and inset zoom views, with image (b) treated as the reference.}
 \label{fig:asteroidImpact}
\end{figure*}
\vspace{-3.2mm}
\fix{\subsection{Analysis of Simulation Datasets with Parametric Vs. Nonparametric Models of Uncertainty}}\label{sec:simulation}
\fix{We demonstrate an application of our proposed parametric and nonparametric statistical techniques (\autoref{sec:parametric} and \autoref{sec:nonparametric}) in the context of oceanology for the analysis of ocean eddies. The knowledge of eddy positions is important for oceanologists to understand the transport of energy and biogeochemical particles in oceans.~\autoref{fig:vorticityVisRedSea} visualizes the results of our probabilistic analysis of fibers that represent vortical features of the Red Sea dataset~\cite{TA:2020:redSea}. The dataset was downloaded from the IEEE SciVis Contest 2020 website (\url{https://kaust-vislab.github.io/SciVis2020/}) and comprises ensembles of multiple variables pertinent to the oceanology simulated for $60$ time steps. In this experiment, we visualize the positional uncertainty of fibers corresponding to eddy features of the Red Sea by analyzing $20$ ensemble members of the velocity field for a time step of $40$ over the Gulf of Aden sampled on a grid of resolution $250 \times 200 \times 50$.}

\fix{Initially, we derive the vorticity field for each ensemble member by computing the curl of the velocity field. We then compute the vorticity magnitude and Z (vertical) component of a vorticity vector per grid vertex for each member. \autoref{fig:vorticityVisRedSea}a visualizes a continuous scatterplot with the mean vorticity magnitude plotted on the vertical axis and the mean Z component of vorticity vector plotted on the horizontal axis. Trait $\mathcal{T}$ (cyan polygon) is selected to understand the fiber positions with relatively high vorticity and a negative value of the Z component of a vorticity vector which denote anticyclonic eddy positions.}

\fix{~\autoref{fig:vorticityVisRedSea}b visualizes the mean-field fiber surface corresponding to trait $\mathcal{T}$. The magenta box in the mean-field visualization indicates a rare occurrence of vortical features.~\autoref{fig:vorticityVisRedSea}c visualizes the results for the independent Gaussian (parametric) noise model (the same assumption as in the paper by Zheng and Sadlo~\cite{TA:Zheng:2021:uncertainScatterPlots}), but for nonrectangular polygonal shape (our contributions). The volume rendering for the Gaussian noise model (top image in~\autoref{fig:vorticityVisRedSea}c) indicates the presence of a relatively large number of vortical features inside the region enclosed by the magenta box because there are multiple regions with high interior probability (mapped to yellow). The probabilistic segmentation of the interior probability volume with an isosurface for isovalue $0.7$ (the bottom image in~\autoref{fig:vorticityVisRedSea}c) clearly shows the presence of multiple vortical features inside the region enclosed by the magenta box.~\autoref{fig:vorticityVisRedSea}d visualizes results similar to~\autoref{fig:vorticityVisRedSea}c, but for the independent histrogram noise model with $10$ bins. Both volume rendering and probabilistic segmentation for the nonparametric noise model indicate a relatively low probability of vortical features inside the region enclosed by the magenta box. Such inconsistency regarding the presence of vortical features inside the magenta box across the three statistical models indicates the need for further eddy analysis in the same region.}

\fix{We parallelized the implementation of the independent nonparametric histogram models with OpenMP C++ and CUDA using an approach similar to the one for the histogram results in \autoref{fig:accuracyTiming}. The serial computation of interior probabilities required $83.84$ seconds, whereas computations with four and $16$ threads resulted in a significant speed-up with $22.21$ and $6.755$ seconds, respectively, of computational time. The GPU CUDA computations required only $0.203$ seconds.}

\autoref{fig:asteroidImpact} visualizes the results of our proposed statistical techniques for the asteroid impact dataset~\cite{TA:Patchett:2016:VisualizationAsteroidImpact}. For our experiment, we analyze time step $189$ of a simulation, in which an asteroid $250$ m in diameter bursts at an elevation of 5 km above sea level and impacts the sea at an angle of 45$^{\circ}$ (simulation series YB31 on the IEEE SciVis 2018 contest website (\url{https://sciviscontest2018.org/}). We treat the dataset with resolution $460 \times 280 \times 240$ as the reference. \autoref{fig:asteroidImpact}a visualizes a continuous scatterplot of the temperature and volume fraction of water variables for the reference dataset. \autoref{fig:asteroidImpact}b visualizes the ground truth fiber surface extracted from the reference volume for trait $\mathcal{T}$, as indicated by the dark blue polygon in \autoref{fig:asteroidImpact}a. Trait $\mathcal{T}$ denotes the positions with a relatively high volume fraction of water and temperature. Next, we partition the data into $2\times2\times2$ bricks and summarize each brick with a probability distribution similar to the hixel technique proposed by Thompson et al.~\cite{TA:Thompson:2011:hixelVis}.
 
\autoref{fig:asteroidImpact}c visualizes the mean-field, which results in a significant loss of features caused by data variations in each brick. Figs.~\ref{fig:asteroidImpact}d-\ref{fig:asteroidImpact}f visualize the interior probabilities derived assuming the independent Epanechnikov (parametric) distribution (\autoref{sec:parametric}), independent histograms with two bins per variable (\autoref{sec:nonparametricIndependent}), and correlated histograms with two bins per variable (\autoref{sec:nonparametricCorrelated}), respectively. Note that for the correlated 2D histograms in \autoref{fig:asteroidImpact}f, we compute interior probabilities in closed form (without requiring numerical integration), as described in \autoref{sec:nonparametricCorrelated}. The probabilistic feature recovery with our proposed statistical frameworks is quite remarkable compared to the mean-field features, as illustrated by the boxes in \autoref{fig:asteroidImpact}, with \autoref{fig:asteroidImpact}b as the reference. The nonparametric statistics in Figs.~\ref{fig:asteroidImpact}e--\ref{fig:asteroidImpact}f show a slightly improved recovery of probabilistic features than the parametric statistics in Fig.~\ref{fig:asteroidImpact}d, as anticipated based on our synthetic data experiments (e.g., \autoref{fig:errorAnalysisTangleSphereIntersection}). A relatively small number of noise samples (here, eight per vertex) may not effectively capture noise correlation in the case of correlated nonparamertic statistics (e.g., \autoref{fig:correlatedErrorRate}), thus yielding quite similar results in Figs.~\ref{fig:asteroidImpact}e--\ref{fig:asteroidImpact}f. 
\vspace{3mm}
\section{Conclusion and Future Work}\label{sec:conclusion}
We \fix{extend the prior work by Zheng and Sadlo~\cite{TA:Zheng:2021:uncertainScatterPlots} to} study the positional uncertainty of fibers for uncertain bivariate data. Specifically, we present a closed-form statistical framework for spatial uncertainty quantification of fibers when data noise is characterized by independent parametric and nonparametric probability distributions. We perform our analysis for the uniform, Epanechnikov, and Gaussian kernels. Additionally, we present a numerical integration approach for uncertainty quantification of fibers when noise in bivariate data is correlated. We present our statistical frameworks for arbitrary shapes of FSCP as opposed to being limited to rectangular FSCP~\cite{TA:Zheng:2021:uncertainScatterPlots, TA:Sane2021FCLS}. We show that the nonparametric statistics improve the accuracy of uncertainty quantification and visualization when compared to the parametric and mean-field statistics via experiments on synthetic and simulation datasets.

In the future, we would like to extend our research to multivariate data with more than two variables. For bivariate data with correlated noise assumption, we would like to derive closed-form solutions (instead of numerical integration) similar to the independent noise assumption, if computable. Finally, we are interested in investigating the effect of uncertainty in multivariate data on the topology of fiber surfaces. 

\acknowledgments{
 This work was partially supported by the Scientific Discovery through Advanced Computing (SciDAC) program in the U.S. Department of Energy, the Intel Graphics and Visualization Institutes of XeLLENCE, the Intel OneAPI CoE, the NIH under award number R24 GM136986, the DOE under grant number DE-FE0031880, and the Utah Office of Energy Development. We wish to thank Dr. Jieyang Chen at the Oak Ridge National Laboratory for helping us with the GPU code implementation of fiber uncertainty quantification. We would also like to thank the reviewers of this article for their valuable feedback.}

\bibliographystyle{abbrv-doi-hyperref-narrow}

\bibliography{uncertainFiberSurfaces}
\end{document}